\documentclass[12pt]{article} 
\pdfoutput=1
\usepackage{mathrsfs}
\usepackage{amssymb}
\usepackage{verbatim}
\usepackage{graphicx,epsf,rotate} 
\usepackage{cite,color,xspace}
\usepackage{amsmath}
\newcommand{\kt}{\tilde{k}}
\newcommand{\oldstuff}[1]{}

\begin{document}
\thispagestyle{empty}

\begin{center}
{\Large \bf  Radion/Dilaton-Higgs Mixing Phenomenology \\in Light of the LHC} \\
\vspace*{1cm} \renewcommand{\thefootnote}{\fnsymbol{footnote}} { {\sf
   Peter Cox}, {\sf Anibal D. Medina}, {\sf Tirtha Sankar Ray}, {\sf Andrew Spray}} 
\\
\vspace{10pt}  {\small
{\em ARC Centre of Excellence for Particle Physics at the Terascale,
School of Physics, \\University of Melbourne, 
Victoria 3010, Australia}} 
\normalsize
\end{center}

\begin{abstract}
 Motivated by the bulk mixing $\xi R_5 H^{\dagger}H$ between a massive radion and a bulk scalar Higgs in warped extra dimensions, we construct an effective four dimensional action that---via the $AdS/CFT$ correspondence---describes the most general mixing between the only light states in the theory, the dilaton and the Higgs. Due to conformal invariance, once the Higgs scalar is localized in the bulk of the extra-dimension the coupling between the dilaton and the Higgs kinetic term vanishes, implying a suppressed coupling between the dilaton and massive gauge bosons. We comment on the implications of the mixing and couplings to Standard Model particles.  Identifying the recently discovered $125$~GeV resonance with the lightest Higgs-like mixed state $\phi_{-}$, we study the phenomenology and constraints for the heaviest radion-like state $\phi_{+}$. In particular we find that in the small mixing scenario with a radion-like state $\phi_{+}$ in the mass range $[150,250]$~GeV,  the diphoton channel can provide 
the best chance of discovery at the LHC  if the collaborations extend their searches into this energy range. 
\end{abstract}

\setcounter{footnote}{0}
\renewcommand{\thefootnote}{\arabic{footnote}}
\newpage
\setcounter{page}{1}

\section{Introduction}

Among the most popular models that extend the Standard Model (SM) of particle physics and solve the gauge hierarchy problem are warped extra dimensions \cite{Randall:1999ee} and composite scenarios \cite{Kaplan:1983fs,Georgi:1984af} where the Higgs is identified as a pseudo-Nambu-Goldstone boson (PNGB) of a broken shift symmetry \cite{Agashe:2004rs}.  In the case of warped extra dimensions, the non-factorizable geometry that leads to a slice of $AdS_5$ space is responsible for effectively reducing quadratic contributions to the Higgs mass once the Higgs five-dimensional (5D) scalar  is localized on (brane Higgs) or near (bulk Higgs) the infra-red (IR) brane. Thus, even though there is no implicit symmetry that leads to a light Higgs, once we assume that such a light mass is generated, it remains natural. On the other hand,  explicit calculable realizations of composite PNGB Higgs  were first found in warped scenarios in what are known as Gauge-Higgs Unification (GHU) models.  Here the SM gauge group is 
enlarged to a gauge group G in the bulk of the extra dimension and broken via boundary conditions to the subgroups H (IR-brane) and $SU(2)_L\times U(1)_Y$ (UV-brane) on the branes. In this way, the fifth component of the gauge field $A_5^{\hat{a}}$ that belongs to the coset group $G/H$ has the right quantum numbers to be the Higgs.  Though it is protected by the gauge symmetry at tree-level, it acquires a potential  at loop level that successfully leads to electroweak symmetry breaking (EWSB) and provides a light Higgs mass protected from the UV-physics~\cite{Agashe:2004rs,Medina:2007hz}. Due to the $AdS/CFT$ correspondence and through the language of holography, it was realized that these kinds of models are particular realizations of composite Higgs scenarios where the Higgs is a PNGB arising from the spontaneous breaking of a global shift symmetry G, and where SM particles have a degree of compositeness determined by their coupling to operators that reside in the conformal sector. 

In both realizations, the conformal sector is spontaneously broken and a corresponding Goldstone mode is expected in the theory. In the 5D picture, this mode is known as the radion and is associated with the spin-0 fluctuations of the metric. In order to stabilize the extra dimension the radion is coupled to an additional scalar field~\cite{Goldberger:1999uk}; the gravity-scalar system can provide a stabilizing potential and a mass for the radion, which is expected to be light. From the 4D point of view this can be accomplished if the corresponding Goldstone mode, the dilaton, couples to a nearly marginal operator of dimension  $|\Delta_{\mathcal{O}}-4|\ll 1$~\cite{Rattazzi:2000hs,Goldberger:2007zk}. Since, besides the other well-known particles of the SM, the LHC has recently discovered what seems to be a light scalar state, it seems reasonable to take the approach that the only new light states accessible at the moment at the LHC are the Higgs and the radion/dilaton.  Given that these two light scalar 
states posses the same quantum numbers, mixing between them is expected, which can have important consequences in the phenomenology of this effective two scalar system. 

In this work we begin by studying the less known case of a bulk scalar Higgs in a warped extra dimension that mixes via a term $\xi R_5 H^{\dagger}H$ with the radion. We show how in this way one can arrive at an effective Lagrangian that describes the different mixing possibilities encountered in these kinds of models.  We also show that moving the Higgs from the brane into the bulk of the extra dimension can already have important consequences on how the radion couples to SM particles, leading to a different radion phenomenology compared to the brane Higgs case~\cite{Goldberger:1999un, Giudice:2000av,Csaki:2000zn, Dominici:2002jv, Rizzo:2002pq}. In particular we show that due to the geometry/conformal symmetry of the radion in the bulk, its couplings to 4D scalar kinetic terms vanish, and therefore the radion coupling to massive gauge bosons is suppressed. The 5D construction is used as a tool to obtain the dependence of the radion and Higgs mixing and couplings on the masses $m_h$, $m_r$ and energy scales 
$v_{ew}\equiv 246$~GeV, radion decay constant $\Lambda_r$ which is taken to be of the order of the conformal breaking scale $f$,  $\Lambda_r\approx f\approx 1$~TeV; we allow for freedom in the specific numerical values of dimensionless parameters.

Once the mixing is taken into account, we perform a numerical scan over the relevant parameters  satisfying the most recent constraints from the LHC on Higgs physics and exotic resonant searches. By matching the lightest mixed state's mass and signal strengths to those measured at the LHC for the 125~GeV resonance, we are able to predict the branching fractions and cross-sections for the most relevant decays of the heaviest mixed state. Interestingly, we find that in some regions the production of two light mixed states via the decay of the heaviest mixed state can contribute as much as 30$\%$ to the total production cross section. Furthermore we find that in the case of negligible mixing, a light scalar state with mass in the range $m \sim [150, 250]$~GeV with a sizeable cross-section into diphotons is still allowed by LHC constraints, providing a very interesting motivation to look in the diphoton channel at larger invariant masses than is currently done at the LHC by ATLAS and CMS.

The paper is organized as follows. In sections~2 and~3 we introduce the radion and a simplified model of a bulk Higgs in warped 5D space and compute the different mixing terms that arise in the presence of the $\xi R_5H^{\dagger}H$ bulk term. In sections~4 and~5 we write the effective Lagrangian describing the mixing between the radion/dilaton and the Higgs and provide the relevant couplings and branching fractions.  Sections~6 and~7 contain the LHC constraints used and the phenomenological study of the heaviest mixed state. Our conclusions are given in section~8.

\section{The Radion}

We are interested in a 5D background that preserves 4D Lorentz symmetry, which can always be written in the form,
\begin{equation}
ds^2=e^{-2A(y)} \eta_{\mu\nu}dx^{\mu}dx^{\nu}-dy^2\label{RS}\,,
\end{equation}
where $y$ is the extra dimensional coordinate and $e^{-A(y)}$ a convex function of $y$. In the Randall-Sundrum solution, $A(y)=k y$ where $k$ is the curvature scale, and the space reduces to a slice of $AdS_5$ with boundaries at $y=0$ (UV-brane) and $y=L$ (IR-brane). By an appropriate gauge choice, one can decouple the spin-0 (radion) from the spin-2 (graviton) fluctuations of the metric Eq.~(\ref{RS}). The spin-0 fluctuations are given by,
\begin{equation}
ds^2=e^{-2A(y)-2F(x,y)} \eta_{\mu\nu}dx^{\mu}dx^{\nu}-(1+2F(x,y))^2dy^2 \,.\label{RS-spin0}
\end{equation}
In the absence of a stabilizing mechanism, the radion is massless and it is simple to check that it consists of a single state with a profile in the extra-dimension given by 
\begin{equation}\label{radionprofile}
F(x,y)=e^{2A(y)} \frac{e^{-kL}}{\sqrt{3}M_p}r(x)\equiv e^{2(A(y)-kL)} \frac{1}{\Lambda_r}r(x),
\end{equation}
where we have used that $M_{P}^2\approx M_5^3/k$, with $M_P$ the 4D reduced Planck mass.

In order to stabilize the extra-dimension, it is customary to introduce an additional scalar in the bulk of the extra dimension with corresponding bulk and brane potentials such that the gravity and scalar sectors mix.  The backreaction of the scalar on the geometry provides a mass for the physical state associated with the radion. This will produce deviations from the pure $AdS_5$ solution for the geometry; however, if the backreaction is not large the deviations tend to be small and the approximate form for the radion profile $F(y)\sim e^{2ky}$ holds~\cite{DeWolfe:1999cp}. We will comment in section \ref{sec:backreaction} on the consequences of the backreaction on the radion-Higgs mixing, which are important once the Higgs is moved to the bulk of the extra dimension.

\section{Radion-Higgs mixing}

Light radion/dilaton phenomenology and mixing with an IR-brane localized Higgs has been studied extensively in the literature~\cite{Giudice:2000av,Csaki:2000zn,Dominici:2002jv,Chacko:2012sy,Chacko:2013dra}.  It has been found that current LHC measurements, in particular of the Higgs mass and signal strengths, already put significant constraints on the parameter space of these models~\cite{Desai:2013pga}.  In this paper we study the consequences of moving the Higgs into the bulk of the extra dimension and mixing it with the gravity sector via a bulk term  $\xi R_5H^{\dagger}H$. Such a bulk mixing term was also considered in \cite{Davoudiasl:2005uu} but in the context of higher curvature Gauss-Bonnet terms.  We motivate an effective 4D low energy action that describes all the possible mixing terms that one may encounter between the two light states in the model, the radion and the Higgs.  We also derive the parametric size of these mixing terms.  In this context let us briefly survey the possible localization 
of the Higgs, what this implies for the radion-Higgs mixing 
in the theory, and the possibility of a Higgs as a PNGB of a shift symmetry.

\subsection{ The brane Higgs scenario}

In this case one can simply write the Higgs part of the Lagrangian as follows:
\begin{equation}
 S_{brane} = \int d^4x \sqrt{|\gamma(r(x))|}\left[ |{\mathcal{D}} H|^2 - V(H) + \xi R_4 H^{\dagger}H \right] \,,
\end{equation}
where $\gamma$ is the induced metric on the boundary. After the Higgs gets a vev $v$, one can perform a Taylor expansion of the potential,
\begin{equation}
 V(H) = \sum \left. \frac{\delta^n V(H)}{\delta H^n}\right|_{H\rightarrow v} h^n.
\end{equation}
The mass mixing term that can arise from the $n=1$ term in the above equation vanishes exactly due to the minimization condition. No mixing of any type arises from the kinetic term, as $\partial_\mu v = 0$.  This is the reason for the absence of any mass mixing in brane Higgs models. Only kinetic mixing via the usual term $\xi R_4 H^{\dagger}H$ is expected.

\subsection{ The bulk Higgs scenario}\label{sec:bulkhiggs}

 Let us now consider a scenario where the Higgs and the SM fields can access the 5D bulk. We use this model as a tool to motivate our effective action in section \ref{sec:effaction} and therefore we briefly describe the process of EWSB and Higgs mass generation.  Technical details of the calculation that are similar to those of~\cite{Cacciapaglia:2006mz} are deferred to appendix~\ref{app:details}.  In this case the full Higgs-radion action may be written as
\begin{align} \label{eq:action}
  S_{bulk} & = \int_0^Ld^5x\,\sqrt{g}\left[\left(\frac{M^3}{2}+{\xi}H^{\dagger}H\right)R_5+{\vert}D_MH\vert^2-V(H)\right] \notag \\
  & \quad -\sum_{\alpha=0,1}{\int}d^4x\,\sqrt{\gamma}\bigg[\left(M^3+2{\xi}H^{\dagger}H\right)[K]+\lambda^\alpha(H)\bigg],
\end{align}
where $V(H)=-6k^2M^3+c^2k^2|H|^2$ is the 5D bulk potential ($c$ a dimensionless localization parameter), $\lambda^{\alpha}(H)$ are the 4D brane potentials, $\gamma$ is the induced metric, and $[K]$ denotes the jump in the extrinsic curvature across the brane.  Note that in adding the direct coupling between the Higgs and the scalar curvature in the bulk, we must also modify the Gibbons-Hawking term to ensure the correct cancellation of boundary terms.  EWSB is induced on the IR brane by taking
\begin{equation}
\lambda^{1}(H)=\frac{1}{2}\frac{\tilde{\lambda}}{k^2}\left(|H|^2-\frac{\tilde{v}^2_{IR}k^3}{2}\right)^2\label{IRpotential},
\end{equation}
where $\tilde{\lambda}$ and $\tilde{v}_{IR}$ are dimensionless quantities. On the UV brane, we simply add a mass term
\begin{equation}
\lambda^{0}(H)=m_{UV}|H|^2.
\label{eq:UVpotential}\end{equation}

To simplify our analysis, we assume that the Higgs back reaction on the metric is negligible.  This requires that the Higgs vev satisfy
\begin{equation} \label{eq:backreaction}
\vert\xi\vert v^2\ll M^3,\qquad \vert v'^2-c^2k^2v^2 + 16 \xi A' v \, v' \vert\ll 12 A'{}^2 M^3 .
\end{equation}
The explicit mixing terms of Eq.~\eqref{eq:action} contribute to the effective bulk and brane masses for the Higgs.  It is straightforward to solve for the Higgs vev $v(y)$.  Expressing it in terms of the physical observable $v_{ew}$, we find
\begin{equation}
v(y)=\sqrt{2(1+\beta)k} \, e^{ky}e^{(1+\beta)k(y-L)}v_{ew}\label{vev},
\end{equation}
where $\beta^2 = 4 + c^2 + 20 \xi$.  The explicit relation between $v_{ew}$ and the 5D parameters is
\begin{equation}
v^2_{ew}\approx\frac{\tilde{\lambda}\tilde{v}^2_{IR} + 16 \xi-2(2+\beta)}{2(1+\beta)\tilde{\lambda}} \,\kt^2 ,\label{v1}
\end{equation}
where $\kt \equiv k e^{-kL}$ and we have neglected terms suppressed by additional powers of $e^{-kL}$.  Inserting our expression for $v(y)$ into Eq.\eqref{eq:backreaction}, we find that the back-reaction is negligible for $O(1)$ values of $\xi$, $\beta$, and $c$, provided that both $k/M<1$ and $v_{ew}/\kt < 1$.

The Higgs fluctuation $h(x, y)={\mathcal{H}}(y)h(x)$, with mass $m_h$, has a more complex form.  In the limit that the Higgs mass is small compared to the RS scale $\kt$, we find that the profile is approximately proportional to the vev:
\begin{equation}
{\mathcal{H}}(y)=\sqrt{2(1+\beta)k} \, e^{ky}e^{(1+\beta)k(y-L)} + \mathcal{O}(m_h^2/\kt^2) .\label{fluctuation}
\end{equation}
Using the IR b.c. one also can determine the mass $m_h$. The resultant equation is complicated in the general case.  However, in our limiting case $m_h \ll \kt$, one can obtain an approximate analytical expression for the lightest mode given by
\begin{equation}
m^2_h\approx 4 (1+ \beta)^2 \tilde{\lambda} v_{ew}^2 . \label{mhiggs}
\end{equation}

To investigate the mass mixing induced by the bulk Lagrangian Eq.~(\ref{eq:action}), we expand the scalar curvature using the $AdS_5$ metric and the replacement $F(x,y)=F(y)r(x)$:
\begin{align}
 R_5 &= [20 A'(y) - 8A''(y)] - 2 e^{2A} F(y) \, \partial^2 r(x) \notag \\
 & \quad + [-80 F(y) A'(y)^2 + 56 A'(y)F'(y) + 32 F(y)A''(y) -8 F''(y)]r(x) + \mathcal{O}\bigl( r(x) \bigr)^2, \notag \\
 \left[K\right] &= 4 A'(y) + \mathcal{O}\bigl( r(x) \bigr)^2. \label{scalarcurvature}
\end{align}
Now using Eq.~\eqref{radionprofile} and $A(y)\sim ky$ we find that the expressions reduce to
\begin{equation} \label{eq:curvkin}
 R_5= 20k^2 - 2 e^{2A} F(y) \, \partial^2 r(x)~~\text{and}~~ [K] = 4k.
\end{equation}
The non-derivative terms linear in the radion vanish.  There could be a residual mass mixing that arises from the product of the constant terms in Eq.~\eqref{eq:curvkin} with the linear fluctuation in the volume element.  However, as discussed below Eq.~\eqref{eq:backreaction}, these constant terms are effective bulk and brane masses for the Higgs, and are more naturally associated with the mixing from the potentials.  Indeed, one can explicitly redefine the Lagrangian mass parameters to absorb these constant terms:
\begin{equation} \label{eq:redefpot}
c^2\rightarrow c^2-20\xi,\qquad
\tilde{v}_{IR}^2\rightarrow\tilde{v}_{IR}^2-\frac{16\xi}{\tilde{\lambda}},\qquad
m_{UV}\rightarrow m_{UV}+8\xi k.
\end{equation}
This will naturally lead to modifications to the definition of $\beta$, and the relation between $v_{ew}$ and the Lagrangian parameters.  Finally we compute the mass mixing that might arise from the potential terms in the bulk and on the brane and the kinetic term in the bulk,
\begin{align} \label{mass_mixing}
-\int d^5 x\sqrt{g} \, c^2 k^2 |H|^2 &\,\to\, 2(\beta^2-4)\frac{v_{ew}}{\Lambda_r}\tilde{k}^2 \, h(x)r(x), && \text{[Bulk potential]} 
\notag \\
-\int d^4x \sqrt{\gamma} \, \lambda^{1}(H) &\,\to\, -8(1+\beta)(2+\beta)\frac{v_{ew}}{\Lambda_r}\tilde{k}^2 \, h(x) r(x) , && \text{[IR brane potential]}
\notag \\
\int d^5 x\sqrt{g} \, g^{55}D_5H^\dagger D_5H &\,\to\, 6(2+\beta)^2\frac{v_{ew}}{\Lambda_r}\tilde{k}^2 \, h(x)r(x). && \text{[Higgs kinetic term]}
\end{align}
We have neglected the UV potential as the Higgs is localized near the IR brane.\footnote{One can show that the contribution from the UV potential cancels with additional exponentially suppressed terms that have been omitted in Eq.~\eqref{mass_mixing}.}  We find that these contributions cancel exactly and leave us with no mass mixing between the Higgs and the radion.

The derivative terms in Eq.~\eqref{eq:curvkin} lead to a kinetic mixing, as in the brane Higgs case.  A quantitative difference from the brane scenario is that the size of the induced mixing is $\beta$-dependent.  Specifically,
\begin{equation}
  \int d^5 x \, \sqrt{g} \, \xi \, R_5 H^\dagger H \to 2 \xi \frac{1+\beta}{2+\beta} \, \frac{v_{ew}}{\Lambda_r} \, (\partial^\mu h(x)) (\partial_\mu r(x)) \, .
\end{equation}
The mixing term also gives contributions to the radion kinetic term.  One contribution arises when the linear derivative term combines with the linear term in $\sqrt{g}$.  Another contribution of the same order comes from the terms in $R_5$ quadratic in the radion.  The net result is
\begin{equation}
  \int d^5 x \, \sqrt{g} \, \xi \, R_5 H^\dagger H \to 3 \xi \frac{1+\beta}{3+\beta} \, \frac{v_{ew}^2}{\Lambda_r^2} \, (\partial^\mu r(x)) (\partial_\mu r(x)) \,. 
\end{equation}

\subsection{Bulk Higgs with back reaction}\label{sec:backreaction}

In the above discussion we did not consider the back reaction of the Higgs and the radion on the metric.  This will modify the bulk profiles of the Higgs, the Higgs vev and the radion. The Higgs back reaction can be assumed to be small as already argued; even if we include its effect, it can at most induce a mass mixing proportional to the Higgs mass $ \sim m_h^2 v_{ew}/\Lambda_r$.  A mass mixing of this order can also arise if we include the differences between the Higgs and Higgs vev bulk profiles $v(y) \neq {\mathcal{H}}(y)$; that is, if we expand the Bessel functions in Eq.~\eqref{eq:bessexp} to include sub-leading terms in $\epsilon$. 

A larger contribution to the mass mixing may arise due to the back reaction of the radion.  Let us assume that the radius stabilizing mechanism results in a small perturbation in the bulk profile of the fields. We can write the following ansatz for the perturbed radion bulk profile and the metric:
\begin{align}
 F(y)&\sim  N_r e^{2ky}(1+l^2 f(y)) ,\notag \\
 A(y) &\sim  ky + \frac{l^2}{6}e^{-2uy},
\end{align}
where $uL = \phi_T/\phi_P$;  $\phi_{T(P)}$ is the radion vev on the TeV (Planck) brane introduced to stabilize the bulk; and $l^2=\phi_P/\sqrt{2M^{3}}.$ The equation of motion for the radion field can be solved using the above ansatz as an expansion in $u/k=\epsilon \sim 1/37$~\cite{Csaki:2000zn}.  Expanding up to $\epsilon^2$ we obtain
\begin{equation}
  f(y) = \frac{1}{3} \left( (1 - \epsilon)e^{-2uy} + \epsilon^2 \left( e^{-2ky} - e^{2k(y-L)} \right) \right)  \,.
\end{equation}
One can now solve for the normalization factor for the radion at this order of the expansion,
\begin{equation}
 N_r = \frac{1}{\Lambda_r} \left( 1+ e^{-2uL} \frac{l^2}{6} (-1+\epsilon + 2\epsilon^2)\right) .
\end{equation}
Solving for the Higgs vev in the same approximation yields,
\begin{equation}
 v(y) = B e^{(2+\beta)ky}\left( 1 + \frac{l^2}{3} \, \frac{2+\beta}{\beta} \, \biggl( 1+ \frac{\epsilon}{\beta} + \frac{\epsilon^2}{\beta^2} \biggr)e^{-2\epsilon ky}\right),
\end{equation}
where $B$ is given by,
\begin{eqnarray}
 B^2&=& 2kv_{ew}^2(1+\beta)e^{-2(1+\beta)kL}\nonumber\\
 &\times&\left(1+ \frac{l^2}{3\beta^3}\left( -(4+\beta)\beta^2 + \frac{(\beta(\beta + 8) +4)\beta}{\beta +1}
 \epsilon - \frac{\beta(5\beta^3 +18\beta +14)+4}{(\beta + 1)^2} \epsilon^2\right) \right) .
\end{eqnarray}
Finally we will assume that,
\begin{equation}
 h(x,y) = h(x)\left( \frac{v(y)}{v_{ew}} + {\mathcal{O}}(m_h^2) \right)
\end{equation}
Thus including the radion back reaction, the Higgs-radion mixing action at leading order can now be written as,
\begin{equation}
 S_{h-r} = \int d^4x\left[ -\xi \frac{2(1+\beta)}{2+\beta}\frac{v_{ew}}{\Lambda_r} h(x) \partial_{\mu}
 \partial^{\mu}r(x) + \left(\xi \frac{2(7+4\beta)}{2+\beta} 
 -\frac{2(1+\beta)}{\beta} \right)\frac{v_{ew}}{\Lambda_r}m_r^2 h(x) r(x)   \right],\label{massmix}
\end{equation}
where $m_r$ is the radion mass given by\cite{Csaki:2000zn},
\begin{equation}
 m_r^2 = \frac{8}{3} l^2 (k\epsilon)^2e^{-2kL}. 
\end{equation}
As expected we find that the mass mixing terms arising from the radion back reaction are proportional to the radion mass.

\subsection{Composite Higgs models}

As mentioned in the introduction, the 5D analogue of the PNGB composite Higgs is the GHU scenario where the Higgs is identified as the fifth component of a 5D gauge boson  $A_M=(A_\mu,A_5)$ belonging to the coset group $G/H$. The higher-dimensional gauge symmetry translates to a 4D shift symmetry of the Higgs.  In a slice of $AdS_5$, the $A_5$ sector of the gauge boson kinetic term of the bulk Lagrangian is
\begin{eqnarray}
    &-&\frac{1}{2}\int d^4 x\, dy~e^{-2ky} \left[(\partial A_5)^2 -2
       \eta^{\mu\nu}\partial_\mu A_5 \partial_5 A_\nu \right] + \dots~ ~~\underrightarrow{\mbox{gauge-fixing}}\nonumber\\
       &-&\frac{1}{2}\int d^4 x\, dy~e^{+2ky} (\partial_\mu A_5^{(0)}(x))^2 
     + \dots~.
\end{eqnarray}
Notice that the higher-dimensional gauge symmetry prevents a tree-level mass for $A_5$ both in the bulk and on the brane.  Also, the antisymmetric nature of the field strength tensor prevents a term like $|\partial_y A_5(y,x)|^2$.  This immediately implies that a composite Higgs cannot have mass mixing with the radion even when the back reaction is considered.  The only possible mixing can be introduced on the brane after the shift symmetry is explicitly broken by the Yukawa and SM gauge interactions to develop a potential. The relevant brane term reads,
\begin{equation}
 \left. S_{CH} \right|_{h-r} = \int d^4x \, \xi_4 R_4 H^{\dagger} H. 
\end{equation}
One can estimate the size of $\xi_4$ by noticing that PNGB potentials are generated at loop level primarily through the top Yukawa which is also responsible for the Higgs developing a potential. Naive dimensional analysis suggests that 
\begin{equation}
\xi_4 \sim \frac{m_h^2}{f^2}\,\Gamma(v_{ew}/f),
\end{equation}
where $f$ is the compositeness scale given by $ke^{-kL}$ and $\Gamma(v_{ew}/f)$ is a generic function of $v_{ew}/f$. Thus, we expect the kinetic mixing induced by this term to be very small.

\section{Effective action}\label{sec:effaction}

Up to this stage we have worked with a 5D warped scenario, considering two particular examples of EWSB, in both of which the Higgs resides in the bulk of the extra dimension. In this way we have been able to determine the possible induced mixing terms between the radion and the Higgs. Though we have determined these mixing terms for a particular scenario we expect their dependence on physical quantities to be general. In fact, from the 4D point of view through the $AdS/CFT$ correspondence, we are describing a scenario of a conformal sector that is spontaneously broken leading to a light pseudo-Nambu-Goldstone boson known as the dilaton. This light state can mix with the other light state in the theory, the Higgs, via  the conformally covariant generalization of the gauge covariant derivative\cite{Salam:1970qk,Vecchi:2010gj}:
\begin{equation}
\left(D_{\mu}-\Delta \frac{\partial_{\mu} r(x)}{r(x)}\right)H(x)+(1-6\xi)H^{\dagger}D_{\mu}H\frac{\partial^{\mu} r(x)}{r(x)}\,,
\end{equation}
where $\Delta$ is the Higgs conformal weight, $D_{\mu}$ is the gauge covariant derivative and we have included an additional term as suggested in Ref.~\cite{Vecchi:2010gj} in order to account for the breaking of the special conformal symmetries, which in the 5D picture corresponds to the case $\xi\neq 1/6$.  This interaction leads to kinetic mixing as found in the previous section. This mixing is always present due to the remnant shift symmetry of the model, and it is the only possible type of mixing allowed  when the CFT is broken spontaneously, as we also saw in our simplified 5D calculation.  An explicit breaking of the conformal symmetry is signalled by the presence of a non-vanishing dilaton mass and consequently the possibility of a mass mixing term between the dilaton and the Higgs field. This is represented in the 5D picture by the deformation from $AdS_5$ space due to back reaction effects responsible for the stabilization of the extra dimension and thus, for the generation of the radion mass. As we 
saw, this explicit breaking of the conformal symmetry leads to mass mixing between the radion and the Higgs in the 5D picture as described in Eq.~(\ref{massmix}).

It then becomes clear that from a pure 4D perspective, we can represent the most general effective phenomenological Lagrangian describing the light degrees of freedom of a spontaneously broken conformal sector by,
\begin{eqnarray}\label{effact}
{ \mathcal{L}}_{eff}&=& \frac{1}{2}\partial_{\mu} h(x)\partial^{\mu} h(x) - \frac{1}{2} m_h^2h(x)^2 + \frac{1}{2}\left(1+c_2\frac{v_{ew}^2}{\Lambda_r^2}\right)\partial_{\mu} r(x)\partial^{\mu} r(x) \nonumber\\ 
&-& \frac{1}{2} m_r^2r(x)^2 - c_1 \frac{v_{ew}}{\Lambda_r}  \partial_{\mu} h(x)\partial^{\mu} r(x) -c_3 \frac{v_{ew}}{\Lambda_r} m_r^2 h(x)r(x)\,,
\end{eqnarray}
where $c_1$, $c_2$ and $c_3$ are $\mathcal{O}(1)$ numerical coefficients, and we use the terms radion/dilaton interchangeably. From this point onwards we focus on this phenomenological Lagrangian to describe the possible mixing scenarios that may arise:
\begin{enumerate}
 \item The no mass mixing scenario, $c_3=0$. From the 5D point of view, this case corresponds to a pure $AdS_5$ slice where the back reaction on the geometry from the radion potential that stabilizes the extra dimension can be neglected. Strictly speaking, it is not compatible to have a massive radion and no mass mixing unless a tuning of the parameters is involved such that $c_3=0$. From the 4D point of view, this corresponds to no explicit conformal breaking parameter in the dilaton self interactions and thus to a CFT that is not badly broken.
\item The generic scenario where $c_3,c_1 \neq 0$ corresponds from the 5D point of view to considering the leading back reaction contributions of the radion potential and from the 4D point of view to explicit conformal breaking terms in the dilaton potential.
\item The gauge-Higgs unification/pseudo-Nambu-Goldstone composite Higgs 5D/4D scenarios correspond to $c_1\ll 1$ and $c_3 = 0$ when explicit sources of conformal breaking are neglected.
\end{enumerate}
Despite the fact that a brane or a bulk Higgs may enter in the same mixing category,  the phenomenology can very different due to the way in which in the conformal breaking is felt as we will see in the next section. For the study of the radion-Higgs mixing and its effect on both the Higgs and radion phenomenology at colliders we shall consider $c_1,c_2$ and $c_3$  as  free parameters. We will see in the next section that the GHU/PNGB composite Higgs scenario reduces phenomenologically to the case of a brane Higgs with $c_1 \ll 1$, and has therefore been covered by previous radion studies \cite{Desai:2013pga}.  Thus we focus our scans on covering all possible values for $c_1,c_2$ and $c_3$ for a bulk scalar Higgs, which provides phenomenologically distinct signatures with respect to the brane Higgs case.

One can diagonalize the kinetic term in Eq.~\eqref{effact} by  going to a new basis $h=h'+c_1 (v_{ew}/\Lambda_r) r'/Z$ and $r=r'/Z$, where
\begin{equation}
Z^2=1+(c_2+c_1^2)\frac{v_{ew}^2}{\Lambda_r^2} \,.
\end{equation}
This transformation decouples the kinetic mixing but introduces additional mass mixing terms. The mass matrix in the basis $(r',h')$ then takes the form
\begin{equation}
M   
=
\begin{pmatrix}
M_{11}  & M_{12}   \\
M_{12}  & M_{22}
\end{pmatrix} = 
\begin{pmatrix}
  \frac{m_r^2}{Z^2}+\frac{1}{Z^2}\frac{v_{ew}^2}{\Lambda_r^2}(c_1^2 m_h^2+2c_1 c_3 m_r^2) &    \frac{1}{Z}\frac{v_{ew}}{\Lambda_r}(c_1 m_h^2+c_3m_r^2)\\
 \frac{1}{Z}\frac{v_{ew}}{\Lambda_r}(c_1 m_h^2+c_3m_r^2)  & m_h^2
\end{pmatrix}\,.
\end{equation}
The mass eigenbasis is obtained by the orthogonal transformation
\begin{equation}
\begin{pmatrix}
r'  \\
h'
\end{pmatrix}
= \begin{pmatrix}
 U_{r,-} &    U_{r,+}\\
 U_{h,-} & U_{h,+}
\end{pmatrix}
\, \begin{pmatrix}
\phi_{-}  \\
\phi_{+}
\end{pmatrix} \,.
\end{equation}
where $\Delta=\sqrt{(M_{11}-M_{22})^2+4M_{12}^2}$ and
\begin{align}
U_{r,-}&=\frac{M_{11}-M_{22}-\Delta}{\sqrt{(M_{11}-M_{22}-\Delta)^2+4M_{12}^2}}\,, & U_{r,+} & =\frac{M_{11}-M_{22}+\Delta}{\sqrt{(M_{11}-M_{22}+\Delta)^2+4M_{12}^2}}\,,\nonumber \\
U_{h,-}&=\frac{2M_{12}}{\sqrt{(M_{11}-M_{22}-\Delta)^2+4M_{12}^2}}\,, & U_{h,+} & =\frac{2M_{12}}{\sqrt{(M_{11}-M_{22}+\Delta)^2+4M_{12}^2}}\,.
\end{align}
There are correspondingly two eigenstates; a lighter one $\phi_{-}=U_{r,-} r'+U_{h,-} h'$ and a heavier one $\phi_{+}=U_{r,+} r'+U_{h,+} h'$, with masses:
\begin{equation}
m^2_{\phi_{\pm}}=\frac{1}{2}(M_{11}+M_{22}\pm\Delta)\,.
\end{equation}
The gauge basis is related to the mass basis according to:
\begin{equation} \label{estates}
r=\frac{1}{Z}(U_{r,+}\phi_{+}+U_{r,-}\phi_{-})\,,\qquad h=(U_{h,+}+\frac{c_1}{Z}\frac{v_{ew}}{\Lambda_r}U_{r,+})\phi_{+}+(U_{h,-}+\frac{c_1}{Z}\frac{v_{ew}}{\Lambda_r}U_{r,-})\phi_{-}\,.
\end{equation}

\section{Higgs and Radion couplings, mixing and branching ratios} \label{sec:couplings}

\subsection{Higgs and Radion couplings}

Though we motivated our effective theory by studying a particular 5D scenario, we ultimately focused on an effective 4D picture wherein we consider two types of Higgs sector: i) the Higgs is identified with a light scalar doublet charged under the gauge group $SU(2)_L\times U(1)_Y$; or ii) the Higgs field is identified with a composite PNGB of an enlarged broken global group that contains $SU(2)_L\times U(1)_Y$ as a subgroup. In both cases there is an associated conformal sector that is spontaneously broken at an energy scale $f$ and that in our effective theory translates into the existence of a possible light state; the dilaton. One may be worried about possible contributions to the Higgs couplings arising from mixing or loop-effects involving resonances of the conformal sector.  However, notice that in case i) the only  symmetry additional to those already found in the SM is the spontaneously broken conformal symmetry.  We expect any possible additional composite resonances  besides the dilaton to 
have masses of the order $m_{ress}\sim g_{\rho} f$, with $g_{\rho}\gg 1$ the strong coupling from the conformal sector, making their effects on the Higgs couplings strongly suppressed. In case ii) due to the enlarged global group in which SM particles are embedded and due to the shift symmetry protection of the Higgs, there is a relationship between the Higgs mass and light top fermionic resonances of the form $m_{h}^2\propto m_{t}^2 m_{Q}^2/f$.  Therefore in order to reproduce a light Higgs mass,  one usually finds the existence of light fermionic resonances that couple strongly to the Higgs, with masses $m_{Q}\sim g_{\psi} f\ll g_{\rho} f$.  This can have significant effects, in particular for Higgs couplings to gluons or photons. It has been shown  nonetheless that due to the pseudo-Nambu-Goldstone nature of the Higgs, the resonant fermionic loop contributions  cancel  against the top quark modified Yukawa coupling, and lead to modifications in the coupling to gluons that are suppressed by the ratio $v^2/
f^2\lesssim 0.01$ \cite{Montull:2013mla}.  Therefore, in the two Higgs scenarios considered, we do not expect sizeable deviations of the Higgs couplings from their SM values, and thus for simplicity we restrict the couplings to SM values.\footnote{In the case of generic warped extra dimensional scenarios the mass scale of the lowest lying KK fermions $m_{KK} \lesssim \Lambda_r.$  A naive estimate of the of shift in the Higgs coupling to gluons due to the KK towers of the SM fermions is as follows,
\[ \frac{\delta \Gamma^{KK}_{gg}}{\Gamma^{SM}_{gg}} \sim 4 C(r_{NP})N_{f}\sum_n 
 \frac{v}{m^{(n)}_{KK}}\frac{\partial m^{(n)}_{KK}}{\partial v} \sim {\mathcal{O}(1)} \left(\frac{v}{ke^{-\pi kL}}\right)^2,\]
 where $C(r_{NP})$ is the quadratic Casimir of the KK states and $N_{f}=6$. This translates into a lower limit on the mass of the lightest KK state that may be as large as 3.2~TeV for a $20 \%$ shift in the decay width, which is the resolution of current experimental data. A detailed calculation of this is rather model dependent \cite{Bhattacharyya:2009nb} and beyond the mandate of this paper.} 

Allowing for the possibility of a bulk Higgs implies that some of the known radion couplings to SM fields are modified, in particular those involving radion couplings to the Higgs field itself as well as to massive gauge bosons. We use the 5D language as an easy tool to calculate the couplings and assume a given warp factor $kL$ that solves the hierarchy problem, though our results are general with the replacement $\Lambda_r=f$. As was shown in Ref.~\cite{Csaki:2007ns}, the bulk radion couples at linear order to SM fields through the  bulk stress energy tensor as
\begin{equation}
S_{radion}=\int d^4x dz \sqrt{g} \,F(x,z)\left[\Theta^{M}_{M}-3 g_{zz}\Theta^{zz}\right]\label{rcoupling},
\end{equation}
where the conformal coordinate $z$ is related to the extra-dimensional coordinate $y$ as $dy=e^{-A}dz$, and $\Theta^{MN}$ is the bulk stress energy tensor which can be written as
\begin{equation}
\Theta^{MN}=-\frac{2}{\sqrt{g}}\frac{\delta(\mathcal{L}_{matter}\sqrt{g})}{\delta g_{MN}}=
-2\frac{\delta(\mathcal{L}_{matter})}{\delta g_{MN}}+g^{MN}\mathcal{L}_{matter} \,.
\end{equation}
Focusing on the coupling to SM gauge bosons (massive or massless), using Eq.~(\ref{rcoupling}), one can easily show that due to the bulk kinetic terms for the gauge fields, there will always be a non-vanishing coupling of the form
\begin{equation}
S_{radion}=\int d^4x dy \left(-\frac{1}{2}\right)F(x,y)F_{\mu\nu}F^{\mu\nu}=-\int d^{4}x \frac{r(x)}{4\Lambda_r}\frac{1}{kL}F_{\mu\nu}F^{\mu\nu},
\end{equation}
where we used that in this case $\mathcal{L}_{matter}=-1/4F_{MN}F^{MN}$ and therefore $\Theta^{MN}=-F^{MA}F^N_{A}+\frac{1}{4}g^{MN}F_{AB}F^{AB}$. As was argued in \cite{Csaki:2007ns}, the fact that this tree level coupling is non-vanishing implies that loop effects merely renormalize this tree-level operator. Therefore, loop effects are prominent on the branes where no tree-level coupling is allowed, being stronger on the IR brane where the radion is usually closely localized. This provides the main mechanism of radion production through gluon fusion as is usual in radion scenarios. We refer the reader to Ref.~\cite{Csaki:2007ns} for the appropriate expressions for the radion-digluon and radion-diphoton couplings, including fermion and gauge boson loops as well as QCD and QED trace anomalies respectively.

In addition, via electroweak symmetry breaking (EWSB), there is in principle a possibly large additional coupling of the radion to a pair of massive gauge bosons, which is dominant in the case of a brane-localized Higgs.  As is well-known, the gauge bosons acquire their mass through the kinetic term of the Higgs field, which in the case of a bulk Higgs scalar leads to mass terms for the gauge bosons of the form
\begin{equation}
\mathcal{L}_{matter}=D_{M}H^{\dagger}D^{M}H\to m_{W}^2 W^{+}_{\mu}W^{\mu, -}+\frac{1}{2}m_{Z}^2 Z_{\mu}Z^{\mu}.
\end{equation}
It follows that the contribution to the stress energy tensor is $\Theta^{MN}=-2 D^{M}HD^{N}H^{\dagger}+g^{MN}D_AH^{\dagger}D^{A}H$, which implies that $\Theta^{M}_{M}=3D^{M}HD_{M}H^{\dagger}=3e^{2A}D^{\mu}HD_{\mu}H^{\dagger}$, where the last index is contracted using the Minkowski metric. Now $\Theta^{zz}=g^{zz}D_{A}HD^{A}H^{\dagger}$, and therefore $-3g_{zz}\Theta^{zz}=-3 e^{2A}D^{\mu}HD_{\mu}H^{\dagger}$, which exactly compensates the contribution from $\Theta^{M}_{M}$. Thus the linear radion coupling to the electroweak gauge boson mass terms vanishes in the case of a bulk Higgs. This result can also be checked by simply expanding the metric in its spin-0 fluctuations in the matter action
\begin{align}
\mathcal{S}_{matter}& =\int d^4x dy\sqrt{g}(D_{\mu}H^{\dagger}D^{\mu}H)\notag \\
&=\int d^4x dy \, e^{-4(A(y)+F(x,y))}(1+2F(x,y)) \, e^{2(A(y)+F(x,y))}D_{\mu}H^{\dagger}D^{\mu}H\notag\\
&\approx \int d^4x dy \, (1-4F^2(x,y)+\mathcal{O}(F(x,y)^3)) \, e^{-2A(y)}D_{\mu}H^{\dagger}D^{\mu}H,
\end{align}
where the index on the r.h.s is contracted using the Minkowski metric.  Therefore we also see in this way  that the coupling vanishes. Notice that this result is general for the kinetic term of any scalar. We can understand this result from the 4D point of view as follows: as we just noticed, the vanishing of this particular coupling is geometrical from the 5D point of view. As a matter of fact we can take both the UV and IR branes to infinity, and the results would still hold in pure $AdS_5$-space. In that particular case, it is clear that the conformal symmetry is exact.  If we look at the 4D picture this implies that the 4D-analogue of the radion, the dilaton field, can only couple derivatively to conformally invariant operators, in particular to $D_{\mu}\Phi D^{\mu}\Phi$, where $\Phi$ is a 4D-scalar field. Therefore from Lorentz invariance we see that no linear coupling can be written that derivatively couples the radion to $D_{\mu}\Phi D^{\mu}\Phi$. This has important consequences for the radion 
phenomenology when the Higgs is a scalar in the bulk, since then its coupling to pairs of massive SM gauge bosons only comes from Eq.~(\ref{rcoupling}) and is highly suppressed. 

In the case of gauge-Higgs unification scenarios,  the Higgs field is identified with the fifth component $A_5(x,y)$ of a gauge field belonging to the coset $G/H$ of an enlarged gauge group $G$ that is broken down to the subgroup $H$ via boundary conditions. In that case the equivalent of the scalar kinetic term is given by
\begin{equation}
\mathcal{S}_{matter}=\int d^4x dy \sqrt{g} \, {\rm Tr}[F^a_{\mu 5}F^{a,\mu 5}],
\end{equation}
where the index $a\in G$. Due to the extra index in the kinetic term, there is a non-vanishing radion coupling proportional to the EWSB induced masses
\begin{align}
\mathcal{S}_{matter}&=\int d^4x dy \, e^{-4(A(y)+F(x,y))}(1+2F(x,y)) \, e^{2(A(y)+F(x,y))}\frac{1}{(1+2F(x,y))^2} \, {\rm Tr}[F^a_{\mu 5}F^{a,\mu 5}]\notag\\
&\approx \int d^4x dy \, e^{-2A(y)}(1-4F(x,y)+\mathcal{O}(F(x,y)^2)) \, {\rm Tr}[F^a_{\mu 5}F^{a,\mu 5}],
\end{align}
where the index on the r.h.s is contracted using the Minkowski metric. Thus, in these kinds of scenarios the radion coupling to massive SM gauge bosons is similar to that encountered for a localized Higgs scalar on the IR brane.

Another potential difference with respect to the brane Higgs scenario may arise in the  Yukawa induced SM fermion-radion interactions with the Higgs field which tend to dominate for heavy fermions with respect to other radion-fermion interactions that are momentum suppressed. For that reason we focus on the term
\begin{equation}
\Delta\mathcal{L}_{Y}=-\int d^4x dy \sqrt{g}\, Y_{5}\left[H \bar{f}f+h.c.\right], \label{Yukawa}
\end{equation}
where $Y_{5}$ is the 5D Yukawa coupling. We again expand the spin-0 fluctuations of the metric and use that the left-handed and right-handed fermion well-normalized zero mode profiles are given by
\begin{equation}
f_{L}(y)=\frac{e^{(\frac{1}{2}-c_L)ky}}{N_{L}}\;,\qquad f_{R}(y)=\frac{e^{(\frac{1}{2}+c_R)ky}}{N_{R}},
\end{equation} 
where $f_{L}(y)$ and $f_{R}(y)$ satisfy
\begin{equation}
\int_{0}^{L}dy f^2_{L,R}(y)=1 \qquad \longrightarrow \qquad N_{L,R}=\sqrt{\frac{e^{(1\mp2c_{L,R})kL}-1}{(1\mp2c_{L,R})k}}\,.
\end{equation} 
The upper and lower signs correspond to $N_L$ and $N_R$ respectively, while $c_{L,R}$ are the fermion bulk mass parameters defined by $M_{L,R}=c_{L,R}k$. Using Eq.~(\ref{vev}) for the Higgs vev, we can obtain an expression for the SM fermion masses by integrating the zero-mode profiles for the fermions and the Higgs vev along the extra dimension. In that case we see that we can express the fermion mass as
\begin{eqnarray}
m_f &=&\int_{0}^{L} dy \, e^{-A(y)} \, f_{L}(y)f_{R}(y)v(y)Y_{5}=\nonumber\\
&=& \frac{1}{N_R}\frac{1}{N_{L}}\sqrt{2(1+\beta)k} \, v_{ew}e^{-(1+\beta)kL}\frac{(e^{(2-c_L+c_R+\beta)kL}-1)}{(2-c_L+c_R+\beta)k}Y_5\,. \label{fermionmass}
\end{eqnarray}
The interaction Eq.~(\ref{Yukawa}), once expanded in the spin-0 fluctuation of the metric, takes the form
\begin{align}
-\int d^4x dy \, e^{-A(y)}\, (-2F(y)) &\, Y_5 \, f_{L}(y) \, f_{R}(y) \, v(y) \, (r(x) \bar{f}^0(x) f^0(x)+h.c.)\notag\\
&\approx  \frac{2 m_f}{\Lambda_r}\frac{(2-c_{L}+c_{R}+\beta)}{(4-c_L+c_R+\beta)}\int d^4 x \, (r(x)\bar{f}^0(x) f^0(x)+h.c.)\,,
\end{align}
where in the last line we assume that the fermion and Higgs profiles are IR localized and satisfy $1-c_L+c_R+\beta>0$. So contrary to the gauge-boson case, we notice that the coupling of the radion to, in particular, the top quark can be non-negligible, similar to the case with a localized Higgs field.


Finally we look at the coupling of the radion to two Higgs. For this coupling there is a kinetic mixing contribution coming from $\sqrt{g}R_5 H^{\dagger}H$ as well as contributions from the Higgs kinetic term, bulk Higgs mass $\sqrt{g} c^2 k^2H^{\dagger}H$ and important boundary contributions from the IR-brane potential $\sqrt{g_4}\lambda_{IR}(H)$.  The Higgs kinetic and bulk mass contributions cancel against some of the IR-brane contributions and after replacing $\tilde{\lambda}$ in terms of $m_h^2$ and $v^2_{ew}$ using Eq.~(\ref{mhiggs}), one can write the radion-diHiggs coupling in the form
\begin{equation}
\int d^4 x \frac{1}{\Lambda_r}\left(2m_h^2-\frac{c_1}{2}m_r^2\right)r(x)h(x)^2. \label{radiondiHiggs}
\end{equation}
We have also used the radion equation of motion $\Box r(x)=-m_r^2 r(x)$. Given Eq.~(\ref{radiondiHiggs}), we do not expect large differences arising in comparison with the brane localized Higgs counterpart.

To summarize, after studying the radion couplings to SM particles, we expect the largest modifications in the phenomenology of the bulk scalar Higgs scenario to arise due to the vanishing of the radion-massive diboson coupling proportional to the gauge boson mass. We list for completeness in Table~\ref{couplingsgauge} the most relevant couplings of the unmixed Higgs and radion states, where $\tau_{i,(h,r)}=4 m_i^2/m_{(h,r)}^2$, $F_{1/2}$ and $F_{1}$ are the usual integrals over fermion and gauge boson states running in the loop and  $b_{QED}=-11/3$ and $b_{QCD}=7$ are the $\beta$-function coefficients.
\begin{table}[h]
\begin{center}
\begin{tabular}{|c|c|c|}
\hline
\rule{0mm}{4mm}
   & $h(x)$ & $r(x)$\\[0.3em]
\hline
\rule{0mm}{5mm}
$f\bar{f}$ & $-\frac{m_f}{v}$ & $\frac{m_f}{\Lambda_r}$\\ [0.3em]
\hline
\rule{0mm}{5mm}
$WW$ & $\frac{2m_W^2}{v}$ & $-\frac{2}{\Lambda_r}\frac{1}{kL}$\\ [0.3em]
\hline
\rule{0mm}{5mm}
$ZZ$ & $\frac{m_Z^2}{v}$ & $-\frac{1}{\Lambda_r}\frac{1}{kL}$\\ [0.3em]
\hline
\rule{0mm}{5mm}
$\gamma\gamma$ & $\frac{1}{v}\left(F_{1}(\tau_{W,h})+\frac{4}{3}F_{1/2}(\tau_{t,h})\right)\frac{\alpha_{EM}}{2\pi}$ &  $-\frac{1}{\Lambda_r}\left(\frac{1}{kL}+\left[b_{QED}-F_{1}(\tau_{W,r})-\frac{4}{3}F_{1/2}(\tau_{t,r})\right]\frac{\alpha_{EM}}{2\pi}\right)$\\ [0.3em]
\hline
\rule{0mm}{5mm}
$gg$ & $\frac{1}{v}\frac{\alpha_{3}}{4\pi}F_{1/2}(\tau_{t,h})$ &  $-\frac{1}{\Lambda_r}\left(\frac{1}{kL}+\left[b_{QCD}-\frac{1}{2}F_{1/2}(\tau_{t,r})\right]\frac{\alpha_{3}}{2\pi}\right)$\\ [0.3em]
\hline
\end{tabular}
\caption{Phenomenologically relevant couplings of the gauge states $h(x)$ and $r(x)$ to SM particles.}
\end{center}
\label{couplingsgauge}
\end{table}

\subsection{ Mixing and branching ratios}

Most of the interactions between the radion and SM particles, except those with massive gauge bosons and to the Higgs itself, have the same structure as those of the SM Higgs to fermions and gauge bosons.  So  one can easily obtain most of the decay rates of the mixed states by inspecting the well-known expressions for the Higgs decay rates (see for example \cite{Djouadi:2005gi}) and using the replacements: $m_h \to m_{\phi_{\pm}}$ and  $g_{h}\to g_{\pm}$, where from Eq.~(\ref{estates}),
\begin{equation}
g_{\pm}=\left(U_{h,\pm}+\frac{c_1}{Z}\frac{v_{ew}}{\Lambda_r}U_{r,\pm}\right)g_{h}+\frac{1}{Z}U_{r,\pm}\,g_{r}\,,
\end{equation}
with $g_h$ and $g_r$ the Higgs and radion couplings to SM particles respectively. 

The interactions that have a structure different than those of the Higgs to SM particles are those of the mixed states to massive gauge bosons and among the mixed states themselves. In this case, the decay rate of the mixed states into massive gauge bosons can be written as
\begin{eqnarray}
\Gamma_{\phi_{\pm}WW}&=&\frac{m_{\phi_{\pm}}}{32\pi}\sqrt{1-4\frac{m^2_W}{m^2_{\phi_{\pm}}}}\times\left[\frac{U^2_{r,\pm}}{Z^2}\frac{g^2_{rWW}}{4}m^2_{\phi_{\pm}}\left(1-4\frac{m^2_W}{m^2_{\phi_{\pm}}}+6\frac{m^4_W}{m^4_{\phi_{\pm}}}\right)\right.\nonumber\\
&+&\left. \frac{U_{r,\pm}}{Z}\left(U_{h,\pm}+\frac{c_1}{Z}\frac{v_{ew}}{\Lambda_r}U_{r,\pm}\right)\frac{3}{2}g_{rWW}g_{hWW}\left(1-2\frac{m^2_W}{m^2_{\phi_{\pm}}}\right)\right.\nonumber\\
&+&\left.2\left(U_{h,\pm}+\frac{c_1}{Z}\frac{v_{ew}}{\Lambda_r}U_{r,\pm}\right)^2 \frac{g_{hWW}^2}{4 m^4_W}m^2_{\phi_{\pm}}\left(1-4\frac{m^2_{W}}{m^2_{\phi_{\pm}}}+12\frac{m^4_W}{m^4_{\phi_{\pm}}}\right)\right], \label{WWdecay}
\end{eqnarray}
where $g_{rWW}$ and $g_{hWW}$ are among the couplings listed in Table \ref{couplingsgauge}, and for decays into Z pairs one needs to divide Eq.~(\ref{WWdecay}) by 2, replace $m_W \to m_Z$, $g_{rWW}\to 2 g_{rZZ}$ and $g_{hWW}\to 2 g_{hZZ}$.

Using Eq.~(\ref{radiondiHiggs}) and assuming that $\phi_{+}$ is mostly radion while $\phi_{-}$ is mostly Higgs, as experimental constraints seem to suggest, we can calculate the decay rate of $\phi_{+}$ to a pair of $\phi_{-}$ states,
\begin{eqnarray}
\Gamma_{\phi_{+}\phi_{-}\phi_{-}}&=&\frac{m^3_{\phi_{+}}}{8\pi \Lambda_r^2}\sqrt{1-4\frac{m^2_{\phi_{-}}}{m^2_{\phi_{+}}}}\times\left[\frac{U_{r,+}}{Z}\left(U_{h,-}+\frac{c_1}{Z}\frac{v_{ew}}{\Lambda_r}U_{r,-}\right)^2 \left(2\frac{m^2_h}{m^2_{\phi_{+}}}-\frac{c_1}{2}\right)\right.\nonumber\\
&+&\left.\left(U_{h,+}+\frac{c_1}{Z}\frac{v_{ew}}{\Lambda_r}U_{r,+}\right)\left(U_{h,-}+\frac{c_1}{Z}\frac{v_{ew}}{\Lambda_r}U_{r,-}\right)^2\left(-\frac{m_h^2}{2v_{ew}}\right)\frac{\Lambda_r}{m^2_{\phi_{+}}}\right]^2.
\end{eqnarray}

\begin{figure}[h]
\begin{center}
\includegraphics[height=5cm]{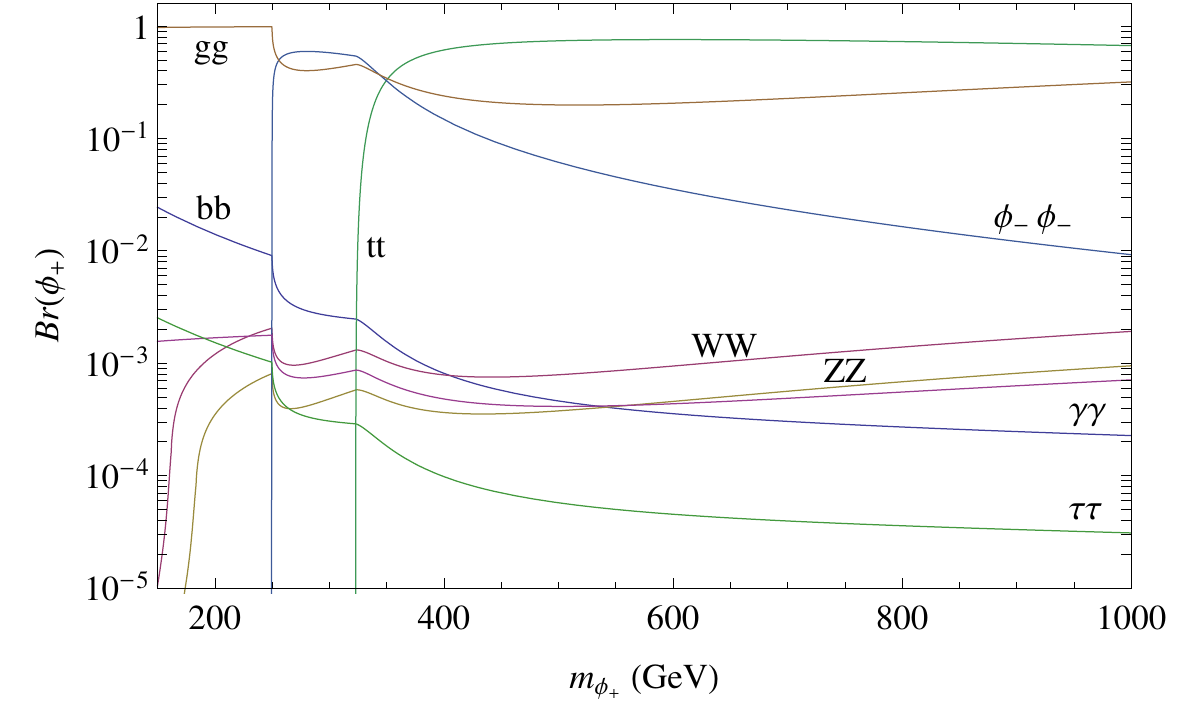}
\end{center}
\caption{Branching ratios for $\phi_+$ as a function of mass with $c_1=c_2=c_3=0$, independent of $\Lambda_r$. }
\label{fig:BR_nomix}
\end{figure}

\begin{figure}[h]
\begin{minipage}[b]{0.5\linewidth}
\begin{center}
\includegraphics[height=5cm]{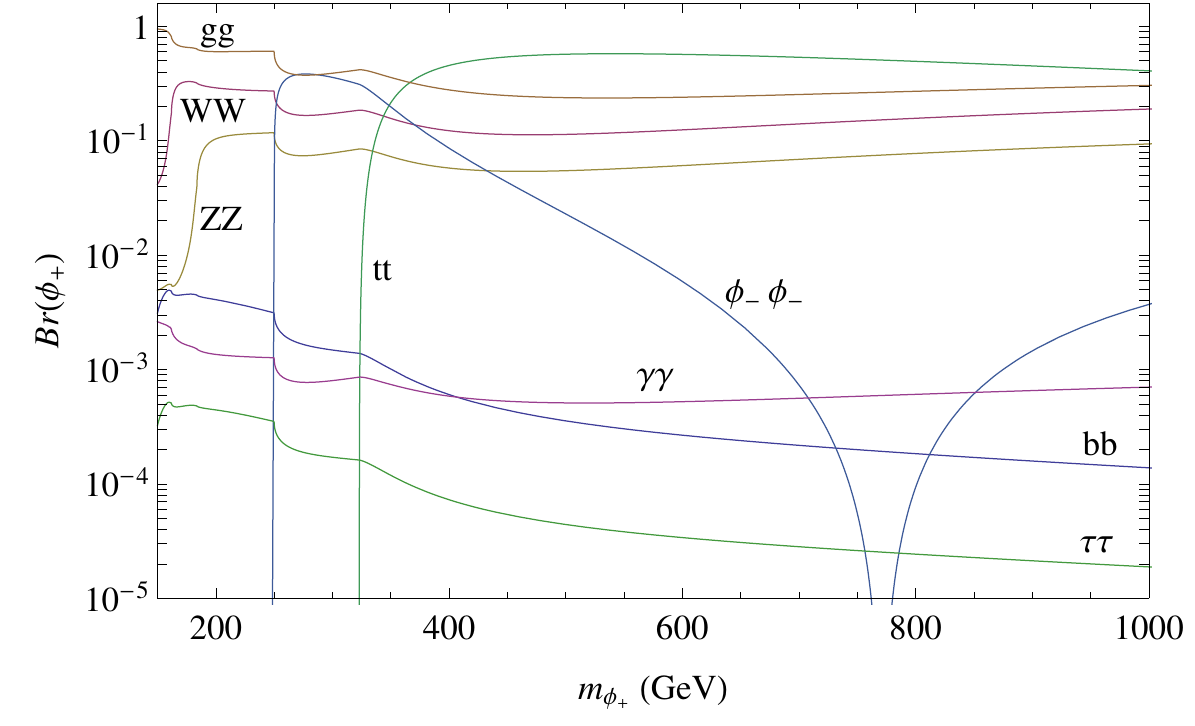}
\end{center}
\end{minipage}
\begin{minipage}[b]{0.5\linewidth}
\begin{center}
\includegraphics[height=5cm]{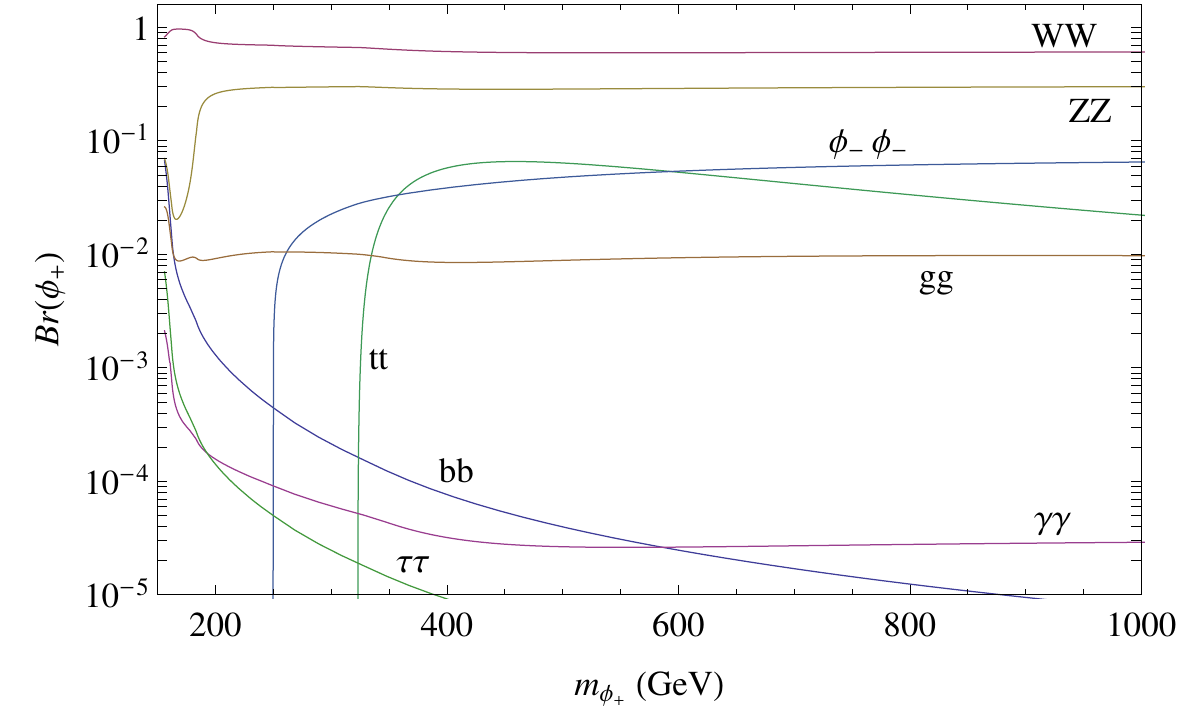}
\end{center}
\end{minipage}
\caption{Branching ratios for $\phi_+$ as a function of mass for $\Lambda_r=3$ TeV. The left and right panels are for $c_1=c_2=c_3=0.1$ and $c_1=c_2=c_3=1$ respectively.}
\label{fig:BRmix}
\end{figure}

 Now that we have all the relevant decay rates, we display in Figs.~\ref{fig:BR_nomix} and \ref{fig:BRmix} the branching fractions of $\phi_+$ as a function of $m_{\phi_{+}}$ for different values of $c_1=c_2=c_3$. In the case of no-mixing ($c_1=c_2=c_3=0$) where $\phi_{+}=r$ and $\phi_{-}=h$, and the results are independent of $\Lambda_r$, notice that for $m_{r}\gg m_{h}$  the dominant decay channels are $t\bar{t}$, $gg$ and $hh$. As already mentioned, decays to massive dibosons only go through their kinetic terms as in the $\gamma\gamma$ channel and tend to be suppressed\footnote{In contrast to the $gg$ channel where the QCD trace anomaly dominates, in the $\gamma\gamma$ channel the anomaly contribution is sub-dominant with respect to the conformal $1/kL$ contribution.}. Thus we expect final state multijets, pairs of b-jets and possibly leptons plus missing energy. Depending on the mass difference between $m_r$ and $m_h$, we could have fat-jets if the decay $h\to b\bar{b}$ is very boosted. For radion masses 
slightly larger than 125 GeV, $m_{r}\gtrsim m_h$, $gg$ dominates with $b\bar{b}$ the second most important decay channel, thus we expect multijets in the final states, which can be hard to differentiate from the SM QCD background found at the LHC at those energies. It is interesting to note that the diphoton channel can have a branching fraction comparable to the SM Higgs for this range of masses and furthermore remains relevant out to higher masses, making it an appealing discovery channel in the small mixing scenario. As shown in Fig.~\ref{fig:BRmix}, once the mixing increases WW and ZZ rapidly become more relevant branching fractions, and at $c_1=c_2=c_3=1$ they dominate over all regions of $m_r > m_h$, relegating the other decay branching fractions to be below 10$\%$.  For smaller mixing, $c_1=c_2=c_3=0.1$, decays into $gg$, $b\bar{b}$ and $\phi_{-}\phi_{-}$ are still relevant, with $gg$ dominant for low $m_{\phi_{+}}$, decays into pairs of $\phi_{-}$ 
important in a small region at intermediate values of $m_{\phi_{+}}$, and $t\bar{t}$ dominating at large values of $m_{\phi_{+}}$. Here we also observe a sharp drop in the branching fraction to $\phi_{-}\phi_{-}$ near $m_{\phi_{+}}\sim750$~GeV due to a cancellation between the various contributions to the partial width. As the mixing is increased, this cancellation occurs at smaller values of $m_{\phi_{+}}$ and eventually moves below the $\phi_{-}\phi_{-}$ threshold and is not observed at $c_1=c_2=c_3=1$.

\section{Constraints from LHC searches}
\label{sec:expconstraints}

Starting with the effective Lagrangian Eq.~\eqref{effact}, we are now in a position to investigate the constraints on the allowed parameter space. With the discovery at the LHC of a new Higgs-like scalar, we consider the case where the lightest eigenstate, $\phi_-$, is identified with this 125 GeV resonance. Measurements of the 125 GeV Higgs signal strengths as well as direct searches for the heavier eigenstate, $\phi_+$, can then be used to constrain the allowed parameter space for models which can be described by Eq.~\eqref{effact}.

The effective Lagrangian we are considering contains six parameters; the mass scales $m_h$, $m_r$, $\Lambda_r$, and the $\mathcal{O}(1)$ dimensionless parameters $c_1$, $c_2$ and $c_3$. One can immediately eliminate $m_h$ by requiring the mass of the lightest eigenstate to be 125 GeV. In addition, we expect $c_2$ to have a very small effect on the phenomenology since the relevant term in the Lagrangian is suppressed by an additional factor of $v_{ew}/\Lambda_r$. Therefore we choose to fix $c_2=1$ before scanning over the remaining 4-dimensional parameter space. We perform a random scan with flat priors over the mass of the radion gauge state, $m_r$, from 160 GeV to 1500 GeV and the kinetic and mass mixing coefficients, $c_1$ and $c_3$, from -3 to 3, while considering fixed values of 1, 3 and 5 TeV for the scale of the radion couplings, $\Lambda_r$. We also note that there is a theoretical bound on $c_1$ in order to ensure that we do not encounter a ghost-like kinetic term for $\phi_+$. For example when $c_
3=0$ and $\Lambda_r=1$ TeV this gives a bound of $\vert c_1 \vert\lesssim4$.

Since we have chosen to identify the lightest eigenstate, $\phi_-$, with the 125 GeV Higgs, we impose the constraints from the measured Higgs signal strengths in the $\gamma\gamma$, $ZZ^{(*)}\rightarrow 4l$, $WW^{(*)}\rightarrow l\nu l\nu$, $b\bar{b}$ and $\tau\bar{\tau}$ decay channels. We focus on the dominant gluon-gluon fusion (ggF) production mode in all channels, with the exception of $b\bar{b}$ where the best measurements are obtained by considering production in association with a W or Z boson (VH). In the case of gluon fusion, the signal strength is defined in the narrow width approximation by
\begin{equation}
\mu^{ggF}_X=\frac{\Gamma(\phi_-\rightarrow gg)}{\Gamma(H_{SM}\rightarrow gg)}\frac{Br(\phi_-\rightarrow X)}{Br(H_{SM}\rightarrow X)}.
\end{equation}
We use the combined ATLAS, CMS and Tevatron best fit values for the signal strengths given in \cite{Belanger:2013xza}, which are shown in Table \ref{sigstrength}, and require that the signal strength for the $\phi_-$ state satisfy these bounds at the 1-sigma level. 
\begin{table}
\begin{center} 
\begin{tabular}[h]{|c|c|} 
\hline
Channel & $\mu_X$ \\ \hline
$\gamma\gamma$ & $0.98\pm 0.28$ \\ \hline
$VV$ & $0.91\pm 0.16$ \\ \hline
$b\bar{b}$ & $0.97\pm 0.38$ \\ \hline
$\tau\bar{\tau}$ & $1.07\pm 0.71$ \\ \hline
\end{tabular}
\caption{Best fit values for the Higgs signal strengths in various decay channels at 125 GeV \cite{Belanger:2013xza}.}\label{sigstrength}
\end{center}
\end{table}

However, for masses $125<m_{\phi_+}<160$ GeV, one must carefully consider the contribution of both states to the measured signal strength in the $WW^{(*)}$ channel, since unlike $\gamma\gamma$ and $ZZ^{(*)}$ the final state is not fully reconstructible. Additionally, interference effects must be taken into account if the two states have a very small mass separation. We therefore restrict $m_r > 160$ GeV, allowing us to consider the two states separately in the $WW$ channel.

We must also consider the possibility that, when kinematically allowed, the 125 GeV state may be produced via the decay $\phi_+\rightarrow\phi_-\phi_-$, which will result in an enhancement in the signal strengths for $\phi_-$. In the case of the $WW^{(*)}$ analysis, such events will not contribute significantly due to vetoes on additional leptons and jets. On the other hand, the $\gamma\gamma$ and $ZZ^{(*)}$ analyses are quite inclusive and this additional contribution to the production cross section can be important. In fact, in certain regions of parameter space this process can become the dominant production mechanism for $\phi_-$. In this case we define the signal strength as
\begin{equation}
\mu^{ggF}_X=\left(\frac{\Gamma(\phi_- \rightarrow gg)}{\Gamma(H_{SM}\rightarrow gg)}+\frac{2\sigma_{ggF}(pp\rightarrow \phi_+ \rightarrow \phi_- \phi_-)}{\sigma_{ggF}(pp\rightarrow H_{SM})}\right)\frac{Br(\phi_- \rightarrow X)}{Br(H_{SM}\rightarrow X)}.
\end{equation}
In addition to the constraints on the 125 GeV eigenstate, the ATLAS and CMS Higgs searches can also be used to constrain the heavier eigenstate. We therefore require that the $\phi_+$ state satisfies the exclusion limits from the CMS $H\rightarrow WW \rightarrow 2l2\nu$ \cite{CMS-PAS-HIG-13-003} and $H \rightarrow ZZ \rightarrow 4l$ \cite{CMS-PAS-HIG-13-002} searches and the ATLAS high mass $H \rightarrow WW \rightarrow e\nu\mu\nu$ search \cite{ATLAS-CONF-2013-067}. The Higgs searches in the remaining channels ($\gamma\gamma$, $b\bar{b}$ and $\tau\bar{\tau}$ ) currently only provide constraints for masses below $\sim150$ GeV. Finally, we also impose the additional constraints provided by the CMS semi-leptonic $t\bar{t}$ resonance search \cite{Chatrchyan:2013lca} in the 500 GeV to 1 TeV mass range.

There are in principle other searches performed at the LHC which could be adapted to our particular model, for example the searches for resonant $ZZ$ production in the dilepton plus dijet channel \cite{CMS-PAS-EXO-12-022} and resonant $WW$ production in the lepton plus dijet channel \cite{CMS-PAS-EXO-12-021}. These searches focus on dibosons produced by KK graviton decay and thus cannot be directly translated to our model without determining the signal acceptance via a Monte Carlo simulation of our signal with appropriate selection cuts. Since this would go beyond the intended scope of the paper, we leave these particular collider studies for subsequent work. While we do not expect these searches to currently constrain our model, they will become important at the 14TeV-LHC. The high mass diphoton \cite{Aad:2012cy} and dijet \cite{CMS-PAS-EXO-12-059} searches may also be able to provide constraints in the future.

\section{Radion-Higgs phenomenology for LHC14}

\begin{figure}[h]
\begin{minipage}[b]{\linewidth}
\begin{center}
\includegraphics[height=6cm]{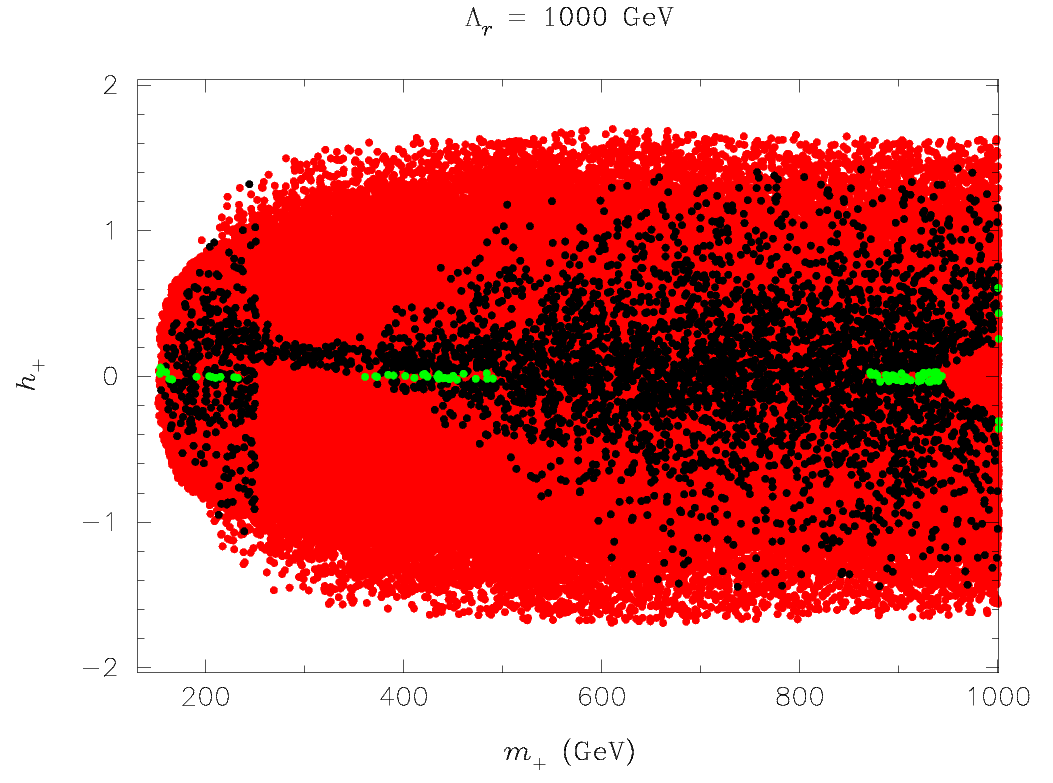}
\end{center}
\end{minipage}
\begin{minipage}[b]{0.5\linewidth}
\begin{center}
\includegraphics[height=6cm]{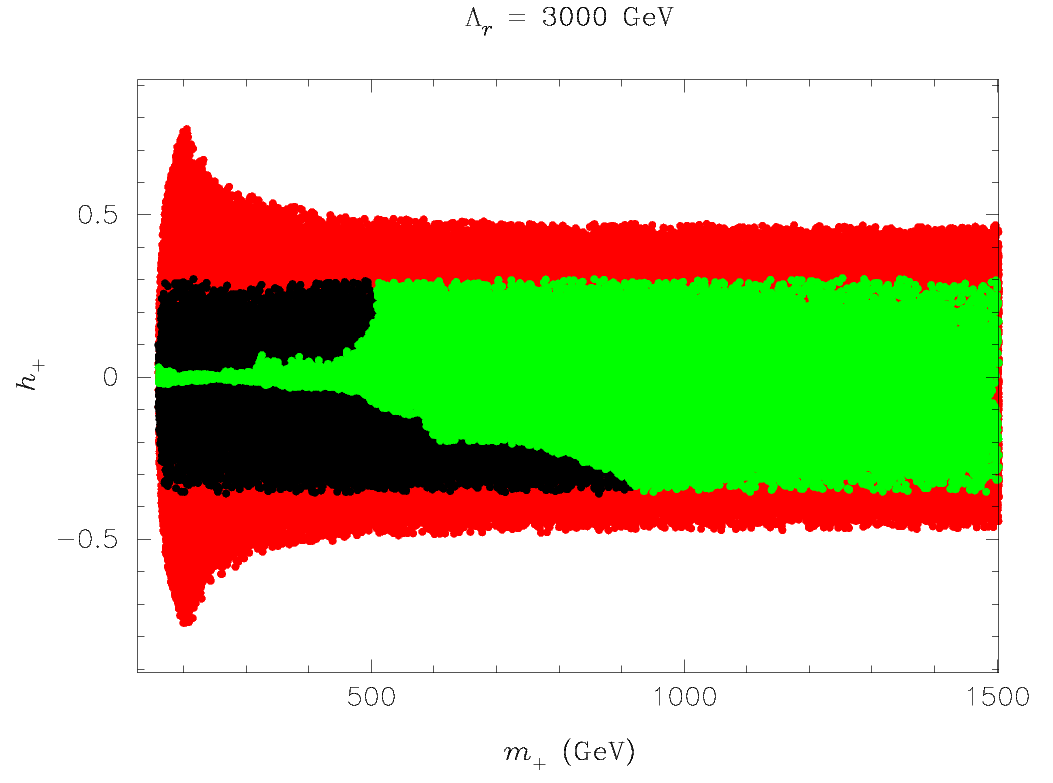}
\end{center}
\end{minipage}
\begin{minipage}[b]{0.5\linewidth}
\begin{center}
\includegraphics[height=6cm]{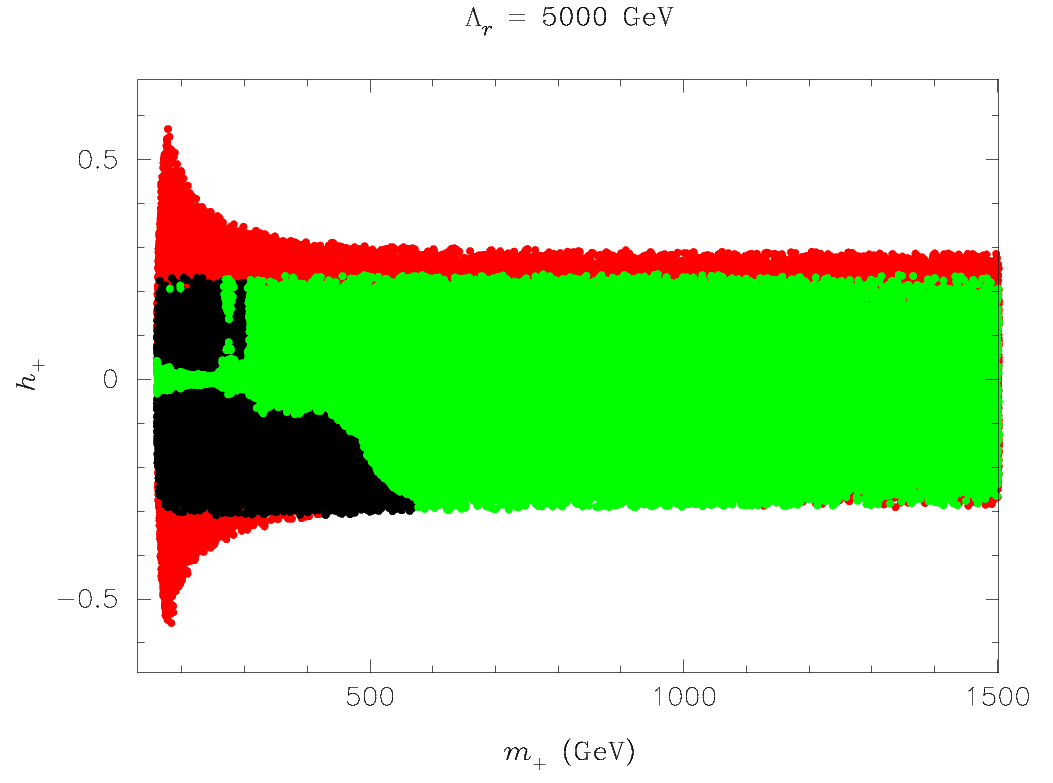}
\end{center}
\end{minipage}
\caption{The fraction of the $\phi_+$ mass eigenstate in the Higgs gauge eigenstate, $h+$, as a function of the mass. The red (dark grey) points are excluded by measurements of the 125 GeV Higgs signal strengths, while the black points satisfy these constraints but are ruled out by direct searches for $\phi_+$. The green (light grey) points pass all of the current experimental bounds. The top panel is for $\Lambda_r=1$~TeV, while the bottom left and right panels are for values of $\Lambda_r=3,5$~TeV respectively.}
\label{fig:h+}
\end{figure}

We performed a scan of 200,000 points for each value of $\Lambda_r$, imposing the above experimental constraints on the $\phi_-$ and $\phi_+$ states. We find that the current experimental constraints already rule out a significant fraction of the parameter space, in particular at low values of $\Lambda_r$ and small masses, $m_+$. Fig.~\ref{fig:h+} shows the fraction of the $\phi_+$ mass eigenstate in the Higgs gauge eigenstate, $h_+=U_{h,+}+\frac{c_1}{Z}\frac{v_{ew}}{\Lambda_r}U_{r,+}$, as a function of the mass. This is a useful variable for characterizing the extent of the mixing between the two states. We note that $h_+$ can be greater than one due to the non-unitary transformation resulting from the kinetic mixing. The red points in Fig.~\ref{fig:h+} are excluded by measurements of the 125 GeV Higgs signal strengths, while the black points satisfy these constraints but are ruled out by direct searches for the heavier state. The green points pass all of the current experimental bounds. The top panel is 
for $\Lambda_r=1$ TeV, while the bottom left and right panels are for values of $\Lambda_r=3,5$~TeV respectively. 

We see that for $\Lambda_r=1$~TeV virtually all of the points are ruled out, with the exception of a few points with very small mixing. In the 250 to 350~GeV range, this is due to enhanced production of the 125~GeV state via $\phi_+\rightarrow\phi_-\phi_-$, as discussed previously. The remaining points which satisfy the Higgs signal strength bounds are excluded by searches for $\phi_+$ in the $WW$ and $ZZ$ channels, as well as the $t\bar{t}$ channel above 500 GeV. In the $\Lambda_r=3$~TeV case we find that, as expected, a significantly larger fraction of the points survive the experimental constraints. Below 500~GeV the bounds from $WW$ and $ZZ$ searches still rule out most of the points with large mixing, while between 450 and 900~GeV we find that they also disfavour negative values of $h+$. This is the result of constructive interference between the Higgs and radion couplings to the top quark, which enhances the $\phi_+$ gluon fusion cross section for negative $h_+$. For masses above 1~TeV there are no 
constraints on $\phi_+$ from current searches. Furthermore, notice that only the red points extend to larger values of $h_+$, which indicates that for large mixing one is unable to satisfy the Higgs signal strength constraints independently of the $\phi_{+}$ mass. Finally, for $\Lambda_r=5$ TeV there are once again significantly more allowed points for masses below 1~TeV, although negative values of $h_+$ are disfavoured by $WW$ and $ZZ$ searches between 250 and 600~GeV.

Using these results we can also derive bounds on the parameters of our effective Lagrangian, in particular $c_1$ and $c_3$. For $\Lambda_r=1$~TeV we find that $-0.2<c_3<0.04$, while the constraints are somewhat weaker for $\Lambda_r=3$~TeV, giving $-2.1<c_3<0.6$. These bounds are of course also dependent on the value of $m_r$ and can be significantly stronger, particularly for lower masses. Considering $c_1$ on the other hand, for $\Lambda_r=1$~TeV we find $-0.2<c_1<0.3$, while for $\Lambda_r=3$~TeV $c_1$ is unconstrained for values of $m_r > 1$~TeV but for masses below 450 GeV we obtain a bound of $-0.7<c_1<2.7$.

\begin{figure}[h]
\begin{center}
\includegraphics[height=7cm]{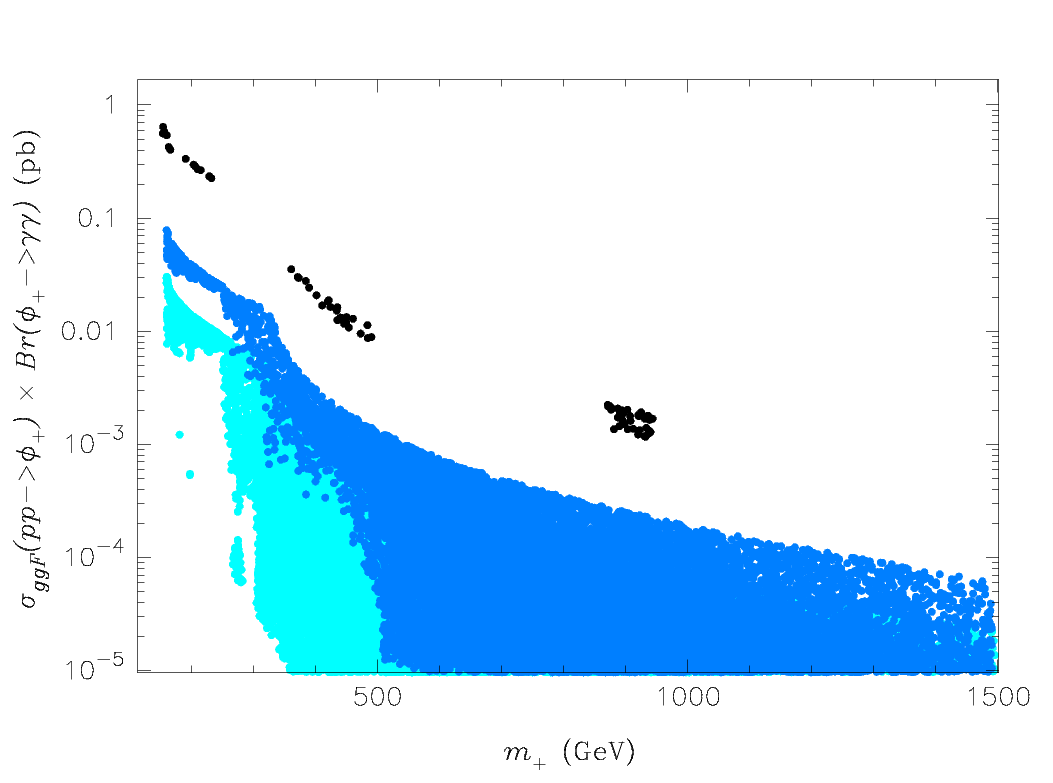}
\end{center}
\caption{Gluon fusion cross section times branching ratio for $\phi_+\rightarrow \gamma\gamma$ as a function of mass. The black, blue (dark grey) and cyan (light grey) points are for $\Lambda_r=$~1, 3 and 5 TeV respectively.}
\label{fig:gamma}
\end{figure}

Finally, we investigate the phenomenology of the regions of parameter space which are allowed by current measurements and discuss the prospects for future searches during the next run of the LHC. We plot in Fig.~\ref{fig:gamma} the diphoton cross-section due to a $\phi_{+}$ produced via gluon fusion at a centre-of-mass energy of 14 TeV, as a function of $m_{\phi_{+}}$. In this and subsequent plots all points satisfy the experimental constraints discussed in section \ref{sec:expconstraints}. The black, blue and cyan points correspond to $\Lambda_r=$~1, 3 and 5 TeV respectively. First of all, notice that the cross-section tends to decrease for larger $m_{\phi_{+}}$, as expected due to the mass suppression in the gluon fusion $\phi_{+}$ production. We concentrate first on the analysis of the $\Lambda_r=1$~TeV (black) points. As mentioned above, the lack of points in the 250--350 GeV mass range can be attributed to the contribution of $\phi_{+}$ decays to $\phi_{-}$ pair-production. This constraint becomes 
suppressed for larger values of $\Lambda_{r}$, and for larger $m_{\phi_{+}}$ due to the reduction in the $\phi_{+}$ production cross section. Similarly, the second empty region is related to the $t\bar{t}$ constraints that kick in at an invariant mass $m_{t\bar{t}}\approx 500$~GeV and which can again be evaded by increasing $\Lambda_r$. 

Another feature that stands out is the relatively large diphoton cross-sections attained for $m_{\phi_{+}}\in [160, 250]$~GeV. Recall that the points in this region correspond to the case of small mixing and therefore the branching ratios of $\phi_{+}$ are dominated by $gg$ and $b\bar{b}$. However, any signal in these channels will be buried under the large QCD background found at the LHC. On the other hand, the clean diphoton signal remains competitive, even overtaking the well-known SM Higgs diphoton cross-section for the same mass range. This can be clearly understood from the fact that $\phi_{+}$ has an enhanced coupling to gluons via the trace anomaly, increasing the production cross section. The diphoton channel is therefore the most promising search channel in this mass range and even extending up to the $t\bar{t}$ threshold since, unlike the SM Higgs, the branching fraction to photons does not drop off at higher masses due to the conformal contribution to the coupling. We therefore strongly encourage 
the CMS and ATLAS collaborations to extend their diphoton searches to invariant masses above the current $m_{\gamma\gamma}=150$~GeV bound. 

\begin{figure}[h]
\begin{center}
\includegraphics[height=7cm]{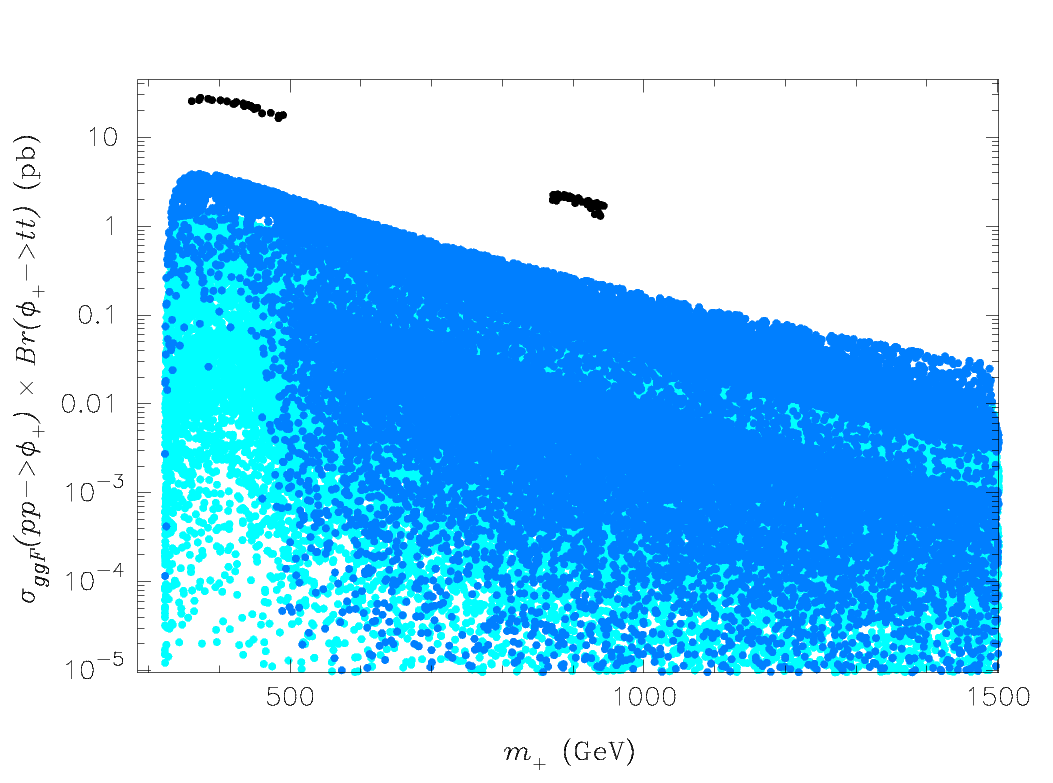}
\end{center}
\caption{Gluon fusion cross section times branching ratio for $\phi_+\rightarrow t\bar{t}$ as a function of mass. The black, blue (dark grey) and cyan (light grey) points are for $\Lambda_r=$~1, 3 and 5 TeV respectively.}
\label{fig:tt}
\end{figure}

Fig.~\ref{fig:tt} shows a similar plot of the cross section for a $\phi_{+}$ produced via gluon fusion and then decaying to $t\bar{t}$, as a function of $m_{\phi_{+}}$. While this search channel currently only provides constraints for $\Lambda_r=1$~TeV, we expect it to become an important channel for masses $m_{\phi_{+}}>500$~GeV with additional integrated luminosity. It provides sensitivity to scenarios with small mixing, where the branching fraction to $t\bar{t}$ dominates. It is particularly sensitive to cases where the mixing parameters in our effective Lagrangian are negative, since this results in an enhanced coupling to $t\bar{t}$ due to the Higgs and radion couplings to the top quark combining constructively. Searches in the $t\bar{t}$ channel will also be important in the high mass region, $m_{t\bar{t}}\gtrsim 1$~TeV, where the decay products are highly boosted and may be collimated into a single jet. Such boosted topologies are already considered by current searches at high invariant mass\cite{
Chatrchyan:2013lca}, although do not currently provide constraints on our model. 

\begin{figure}[h]
\begin{center}
\includegraphics[height=7cm]{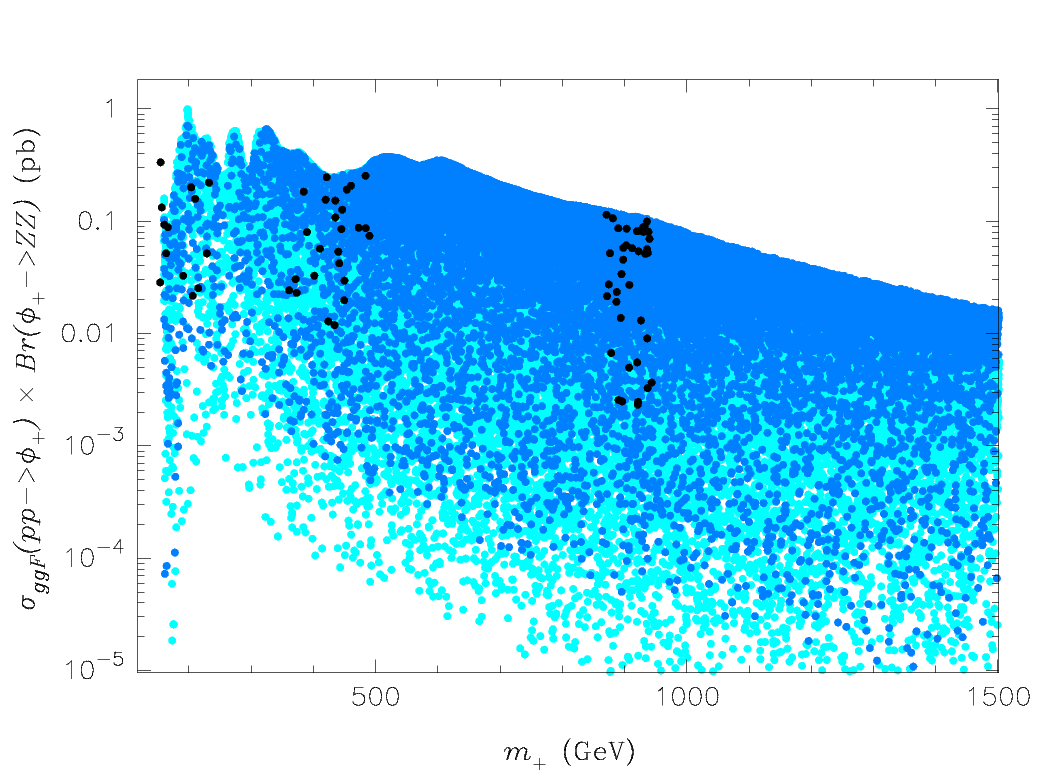}
\end{center}
\caption{Gluon fusion cross section times branching ratio for $\phi_+\rightarrow ZZ$ as a function of mass. The black, blue (dark grey) and cyan (light grey) points are for $\Lambda_r=$~1, 3 and 5 TeV respectively.}
\label{fig:ZZ}
\end{figure}

 Finally, in Fig.~\ref{fig:ZZ} we show the cross section times branching ratio for $\phi_{+}$ decaying to $ZZ$ as a function of $m_{\phi_{+}}$. Again one can clearly see the regions below 1 TeV, and in particular below 500 GeV, where searches in this channel are already restricting the allowed parameter space, even for larger values of $\Lambda_r=5$~TeV. We also note that there are a significant number of points with relatively large cross sections, $\sim 0.1$ pb, for $m_{\phi_{+}}\gtrsim 1$~TeV. These points correspond to cases where there is a large mixing, since as discussed in section~\ref{sec:couplings}, the radion coupling to massive gauge bosons is suppressed when the Higgs is placed in the bulk. On the other hand there are a large number of points with smaller mixing where the cross section is highly suppressed. This ability to suppress the signal in the $ZZ$ (and $WW$) channels provides a distinct difference from the commonly considered brane Higgs scenarios. Hence, while the $ZZ$ channel has 
sensitivity across the entire mass range considered, other channels will be essential to probe the full parameter space. In conclusion we note that the various searches are in fact complementary, with the $ZZ$/$WW$ channels providing the best sensitivity in cases with large mixing, while the $\gamma\gamma$ and $t\bar{t}$ channels are important to probe cases where the mixing is small.

\section{Conclusions}
A light scalar field, associated with the radius stabilization of a compactified extra dimension, is a generic prediction of warped 5D models. In the effective 4D picture this maps to a pseudo-Nambu-Goldstone mode associated with the spontaneous breaking of the conformal symmetry.  The phenomenology of the radion is sensitive to the 5D configuration. In the case of the brane Higgs or the Gauge-Higgs unification scenario, the IR brane assumes a physical significance. In the former picture it is the location of the Higgs, while in the latter it plays a crucial role in breaking of the 5D gauge symmetry through twisted boundary conditions. The existence of the IR brane breaks the conformal symmetry spontaneously, but the presence of a localized Higgs, or the twisted boundary conditions in the GHU case, leads to an explicit breaking of the conformal symmetry. This in turn implies a direct coupling of the radion to the massive gauge bosons which dominates its phenomenology. On the other hand,  we find that in the 
bulk Higgs case,  where the IR brane can be technically pushed to infinity,  the coupling to the massive gauge bosons are suppressed thus providing a significantly different phenomenological scenario.

In this paper we have surveyed 5D scenarios such as the brane Higgs, bulk Higgs and the GHU scenario. Leading contributions from the back reaction of bulk fields that stabilize the extra-dimension were considered. We find that the bulk Higgs scenario provides a distinct and rich phenomenology at colliders. We derived the most generic 4D effective action of the radion/dilaton-Higgs sector for the bulk SM configuration. The relevant couplings of the radion and the bulk Higgs were then computed. The radion coupling to the massive gauge bosons gets suppressed as the radion-Higgs mixing decreases. Thus a relatively unmixed light radion can evade existing experimental searches which are heavily dependent on it decaying to massive gauge bosons. We performed an extensive scan of the parameter space to uncover regions that pass all existing collider bounds with the identification of a light mostly Higgs-like state and a heavier mostly radion-like state.

We find that radion masses as light as 160 GeV are allowed and may have remained hidden in the existing searches. For masses above 250 GeV, decays of the heaviest radion-like state into pairs of light Higgs-like states can contribute to their production by up to 30$\%$. We find that the heaviest mostly radion-like state can be divided into several categories depending on its mass and the extent of the mixing. Below 250 GeV the surviving region corresponds to an almost pure radion-like state with suppressed couplings to massive gauge bosons. The $\gamma\gamma$ channel may be the most sensitive in this region and potentially remain viable at higher masses. Above 500 GeV, both the $t\bar{t}$ and diboson channels will be important at the LHC 14 TeV, in cases of small and large mixing respectively. Beyond $\Lambda_r =1$ TeV scale a large mixing between the radion and the Higgs may be tolerated by the present data. We find that the diphoton, diboson and $t\bar{t}$ channels are complementary and can be used to 
explore large regions of the parameter space. In light of this we urge our experimental colleagues to extend their $\gamma\gamma$ analysis to higher mass scales beyond 150 GeV.

\section*{Acknowledgements}

We would like to thank Tony Gherghetta for helpful discussions and comments on the manuscript, as well as Hiroshi de Sandes for helpful discussions. This work was supported by the Australian Research Council.

\appendix

\section{Higgs VEV and Profiles}\label{app:details}

In this appendix we provide details from the derivation of section~\ref{sec:bulkhiggs}.  Given the action Eq.~\eqref{eq:action}, with brane potentials in Eqs.~\eqref{IRpotential} and~\eqref{eq:UVpotential}, we wish to find the profiles of the Higgs vev $v(y)$ and lightest mode $h(x, y)$.  We begin by expanding the Higgs field in the unitary gauge as a bulk vev plus a fluctuation
\begin{equation}
H(x,y)=\frac{1}{\sqrt{2}}\left(
\begin{array}{c}
0\\
v(y)+h(x,y)\\
\end{array}
\right),
\end{equation}
and concentrate for the moment on the background solution.   Variation of the action Eq.~ \eqref{eq:action} gives the following bulk equations of motion 
\begin{align}
v''-4A'v'+4\xi(2A''-5A'^2)v-\sqrt{2}\frac{\partial V}{\partial H^\dagger}\bigg\vert_{H=\frac{v}{\sqrt{2}}} & =0,\\
A''-\frac{v'^2 + 2\xi ({v'}^2 + v \, v'' + A' v \, v' )}{3(M^3+{\xi}v^2)}&=0,\\
6(M^3+{\xi}v^2)A'^2-\frac{v'^2}{2}+V - 8\xi A' v \, v'&=0,
\end{align}
where primes denote differentiation with respect to $y$.  We also have the boundary conditions
\begin{equation}
A'=\pm\frac{\lambda^\alpha}{3(M^3+{\xi}v^2)},\qquad v'=\pm\left(\sqrt{2}\frac{\partial\lambda^\alpha}{\partial H^\dagger}\bigg\vert_{H=\frac{v}{\sqrt{2}}}-8\xi A'v\right),
\end{equation}
where the upper and lower signs correspond to the UV and IR branes respectively.

We assume that we can neglect the back reaction of the Higgs.  Then we have $A(y)=ky$ and the vev, $v(y)$, is given by
\begin{align} 
v''(y)-4k v'(y)-(c^2+20\xi)k^2v(y)&= 0, \label{eq:VEVEOM} \\
\left(v'(y)+\frac{\tilde{\lambda}}{2 k^2}v(y)\left(v^2(y)-\left(\tilde{v}^2_{IR}+\frac{16\xi}{\tilde{\lambda}}\right)k^3\right)\right)\bigg|_{IR}&= 0,\\
\left(v'(y)-(m_{UV}-8\xi k)v(y)\right)|_{UV}&= 0.
\end{align}
From these expressions we infer the redefinitions noted in Eq.~\eqref{eq:redefpot}. The general solution for the EOM in the bulk then takes the usual form
\begin{equation}
v(y)=A_{1} e^{(2-\beta)ky}+A_{2}e^{(2+\beta)ky},
\end{equation}
where $\beta=\sqrt{4+c^2}$ and $A_1$ and $A_2$ are constants to be determined by the boundary conditions (b.c.).   We use the UV b.c. to select the solution growing towards the IR brane; choosing $m_{UV}=(2+\beta)k$ enforces $A_1 = 0$.  The other constant $A_2$ is fixed by the IR b.c. leading to the solution
\begin{equation}
v(y)=k^{3/2}e^{(2+\beta)k(y-L)}\sqrt{\frac{\tilde{\lambda}\tilde{v}_{IR}^2-2(2+\beta)}{\tilde{\lambda}}}.
\end{equation}
We can relate the constants $\tilde{v}_{IR}$ and $\tilde{\lambda}$ to the electroweak vev $v_{ew}$ by considering the SM gauge boson masses. We must satisfy
\begin{equation}
\int^{L}_{0}dy \, e^{-2ky} \, v^2(y)=v^2_{ew}.
\end{equation}
This directly leads to Eqs.~\eqref{vev} and~\eqref{v1} that we quoted earlier.

We must now check whether this solution does indeed correspond to a small back reaction for the Higgs vev. Evaluating the conditions in  at $y=L$, where $v(y)$ takes its maximum value, we obtain
\begin{align} \label{eq:smallHbr}
  \frac{\vert\xi \vert v^2}{M^3}&=\vert\xi \vert\left(\frac{k}{M}\right)^3\frac{2(1+\beta)v_{ew}^2}{\kt^2}\ll 1, \notag \\
  \frac{\vert v'^2-c^2k^2v^2 + 16 \xi A' v\,v'\vert}{12k^2M^3}&=\frac{1}{12}\left(\frac{k}{M}\right)^3\left((2+\beta)^2-c^2 + 16 \xi (2 + \beta) \right)\frac{2(1+\beta)v_{ew}^2}{\kt^2}\ll 1.
\end{align}
These conditions are easily satisfied for $O(1)$ values of $\xi$, $\beta$, $c$, provided that $k/M<1$ and $v_{ew}< \kt$.

Moving now to the Higgs fluctuation, it satisfies the equations
\begin{align}
{\mathcal{H}}''(y)-4k {\mathcal{H}}'(y)-c^2k^2{\mathcal{H}}(y)+m_h^2e^{2ky}{\mathcal{H}}(y)&= 0, \label{eq:HiggsEOM}\\
\left({\mathcal{H}}'(y)+\left[\frac{\tilde{\lambda}v^2(y)}{k^2}-(2+\beta)k\right]{\mathcal{H}}(y)\right)\bigg|_{IR}&= 0,\\
\left({\mathcal{H}}'(y)-m_{UV}{\mathcal{H}}(y)\right)|_{UV}&= 0,
\end{align}
where in the IR boundary condition we have kept only linear terms in ${\mathcal{H}}(y)$.  Note that Eq.~\eqref{eq:HiggsEOM} differs from Eq.~\eqref{eq:VEVEOM} only through the last term proportional to the Higgs mass, which is a small correction when $m_h \ll \kt$.  So we expect that the Higgs and vev profiles are similar.   The general solution to the bulk equation of motion (EOM) takes the form
\begin{equation}
{\mathcal{H}}(y)=e^{2ky}\left(J_{-\beta}\left(\frac{e^{ky}m_h}{k}\right)\Gamma(1-\beta)B_1+J_{\beta}\left(\frac{e^{ky}m_h}{k}\right)\Gamma(1+\beta)B_2\right),
\end{equation}
where $B_1$ and $B_2$ are constants whose ratio is fixed by the UV b.c and are completely determined once we normalize the 4D kinetic term of the Higgs fluctuation.  Using the UV b.c. and that $\epsilon=m_h/k\ll 1$, we can expand 
the arguments of the Bessel functions
\begin{equation}
J_{-\beta}(\epsilon)=\epsilon^{-\beta}\left(\frac{2^{\beta}}{\Gamma(1-\beta)}+\mathcal{O}(\epsilon)\right), \qquad
J_{\beta}(\epsilon)=\epsilon^{\beta}\left(\frac{2^{-\beta}}{\Gamma(1+\beta)}+\mathcal{O}(\epsilon)\right), 
\label{eq:bessexp}\end{equation}
and find that
\begin{equation}
\frac{B_1}{B_2}\approx \epsilon^{2+2\beta} g(\beta)\label{ratio},
\end{equation}
where $g(\beta)$ is a regular function of $\beta,~$ $g(\beta)\sim \mathcal{O}(1)$ and we have replaced $m_{UV}=(2+\beta)k$. Now at large values of y, the two Bessel functions will behave in an analogous way (neither will be more important than the other in terms of magnitude). Thus at large values of y, given the ratio Eq.~\eqref{ratio}, we see that the solution with $B_2$ dominates. At small values of y this is still the case since the first term in the general solution for ${\mathcal{H}}(y)$ goes as $\epsilon^{2+\beta} B_2$ while the second term goes as $\epsilon^{\beta}B_2$. Therefore, in this limit we can neglect the first term in the general solution for ${\mathcal{H}}(y)$ and write
\begin{equation}
{\mathcal{H}}(y)\approx 2^{-\beta}e^{(2+\beta)ky}\left(\frac{m_h}{k}\right)^\beta B_2,
\end{equation}
where we have used that $m_h\ll k$ and also that $m_h \ll \kt$. Normalizing the 4D kinetic term for the fluctuation according to
\begin{equation}
\int^{L}_{0}dy \, e^{-2ky} \, {\mathcal{H}}(y)^2=1,
\end{equation}
determines the final constant $B_2$ and thus we have Eq.~\eqref{fluctuation},
\begin{equation}
{\mathcal{H}}(y)=\sqrt{2(1+\beta)k} \, e^{ky}e^{(1+\beta)k(y-L)}.
\end{equation}

\end{document}